\documentclass[twocolumn,twoside,10pt,aps,prl,superscriptaddress]{revtex4-2}
\usepackage{amsmath}
\usepackage{makecell}
\usepackage[switch]{lineno}
\usepackage{graphicx}
\usepackage{dcolumn}
\usepackage{bm}
\usepackage[colorlinks=true,linkcolor=blue,citecolor=blue,urlcolor=blue]{hyperref}

\def\arctanh{\operatorname{arctanh}} 
\begin{document}

\title{Quantum Metrology via Floquet-Engineered Two-axis Twisting and Turn Dynamics}

\def\SYSUZH{Laboratory of Quantum Engineering and Quantum Metrology, School of Physics and Astronomy,\\ 
Sun Yat-Sen University (Zhuhai Campus), Zhuhai 519082, China}
\def\SZU{Institute of Quantum Precision Measurement, State Key Laboratory of Radio Frequency Heterogeneous Integration, \\
College of Physics and Optoelectronic Engineering, Shenzhen University, Shenzhen 518060, China}
\def\GD{Quantum Science Center of Guangdong-Hong Kong-Macao Greater Bay Area (Guangdong), Shenzhen 518045, China}

\author{Jihao Ma}
\altaffiliation{These authors contributed equally to this work.}
  \affiliation{\SYSUZH}
  \affiliation{\SZU}

\author{Yi Shen}
\altaffiliation{These authors contributed equally to this work.}
  \affiliation{\SYSUZH}
  \affiliation{\SZU} 

\author{Jiahao Huang}
  \altaffiliation{Email: eqjiahao@gmail.com}
  \affiliation{\SYSUZH}
   \affiliation{\SZU}

\author{Chaohong Lee}
  \altaffiliation{Email: chleecn@szu.edu.cn, chleecn@gmail.com}
  \affiliation{\SZU}
  \affiliation{\GD}
  
\date{\today}

\begin{abstract}
One core of quantum metrology is the utilization of entanglement to enhance measurement precision beyond the standard quantum limit.
Here, we utilize the Floquet-engineered two-axis twisting (TAT) and turn dynamics to generate GHZ-like states for quantum metrology.
Using both analytical semi-classical and quantum approaches, we find that the desired $N$-particle GHZ-like state can be produced in a remarkably short time $t_\mathrm{opt}\propto \ln{N}/{N}$, and its quantum Fisher information $F^\mathrm{opt}_\mathrm{Q}\propto N^2$ approaches the Heisenberg limit.
Owing to the rapid state preparation, it shows outstanding robustness against decoherence.
Moreover, using the Floquet-engineered anti-TAT-and-turn, one may implement an efficient interaction-based readout protocol to extract the signal encoded in this GHZ-like state. 
This Floquet-engineered anti-TAT-and-turn approach offers a viable method to achieve effective time-reversal dynamics to improve measurement precision and resilience against detection noise, all without the need to invert the sign of the nonlinear interaction.
This study paves a way for achieving entanglement-enhanced quantum metrology via rapid generation of GHZ-like states at high particle numbers through continuous Floquet engineering.
\end{abstract}

\maketitle

{\it Introduction.---}
Quantum metrology utilizes quantum entanglement to enhance measurement precision from the standard quantum limit (SQL) to the Heisenberg limit (HL)~\cite{doi:10.1126/science.1097576,doi:10.1126/science.1104149,RevModPhys.89.035002,RevModPhys.90.035005, huang2024entanglementenhanced}, offering promising applications in atomic clocks~\cite{RevModPhys.87.637,pedrozo2020entanglement,10.1063/5.0121372,robinson2024direct}, magnetometers~\cite{PhysRevX.5.031010,thiel2016quantitative,RevModPhys.92.015004,bao2020spin}, gyroscopes~\cite{10.1116/1.5120348,PhysRevApplied.14.034065,Jiao:23,saywell2023enhancing}, gravimeters~\cite{PhysRevLett.117.203003,doi:10.1126/sciadv.aax0800,PhysRevLett.125.100402} and other sensors.
The ultimate measurement precision of an estimated phase $\phi$ is determined by the quantum Cramér-Rao bound (QCRB) $\Delta\phi \geq 1/\sqrt{F_Q}$, defined by the quantum Fisher information (QFI) ${F_Q}$ ~\cite{helstrom1969quantum,FUJIWARA1995119,escher2011general}.
The preparation of an achievable entangled state with high QFI is a top priority. 
Among all classes of entangled states, multi-particle maximally entangled state (i.e., Greenberger-Horne-Zeilinger (GHZ) state)~\cite{doi:10.1126/science.aay0600,doi:10.1126/science.aax9743,cao2024multi,finkelstein2024universal} and GHZ-like states that share similarities with the GHZ state~\cite{PhysRevLett.108.183602,doi:10.1126/science.1250147,PhysRevLett.122.173601,PRXQuantum.2.030204} (e.g., spin cat states~\cite{huang2015quantum,PhysRevA.98.012129,PhysRevA.105.062456}) are among the ultimate goals for both quantum metrology and quantum information~\cite{Zhao2021}.
Although the dynamics of the one-axis twisting (OAT)~\cite{PhysRevA.47.5138,MA201189} can generate these entangled states, it requires a significantly long evolution time for preparation\cite{PhysRevLett.82.1835}, which poses a substantial challenge~\cite{doi:10.1126/science.aar3102}.

It has been demonstrated that OAT-and-turn~\cite{PhysRevA.67.013607,PhysRevA.92.023603} and two-axis twisting (TAT)~\cite{PhysRevA.92.013623,Borregaard_2017,PhysRevLett.129.090403} can achieve faster entanglement generation.
Although TAT exhibits a superior speedup, its interaction form does not naturally exist in known physical systems~\cite{luo2024hamiltonian,miller2024twoaxis}.
In principle, an effective TAT interaction can be realized with OAT-and-turn, in which one can transform OAT into TAT by applying suitable transverse couplings~\cite{PhysRevLett.107.013601,PhysRevA.90.013604,PhysRevA.91.043642,PhysRevA.100.041801}. 
The coupling field can be a train of periodic pulses~\cite{PhysRevLett.107.013601,PhysRevA.90.013604}, a continuous periodic modulated field~\cite{PhysRevA.91.043642}, or even a sequence designed by machine learning~\cite{PhysRevA.100.041801}. 
Besides, by applying Floquet-engineered pulses in OAT, an artificial three-body interactions (named as XYZ model) may be realized to rapidly generate GHZ-like states with high QFI~\cite{PhysRevLett.132.113402}.
A natural question arises: \textit{can we use turn dynamics to accelerate TAT and achieve faster GHZ-like state generation by applying Floquet-engineering?}

On the other hand, extracting the signal encoded in GHZ-like states generally requires single-particle-resolution measurements of parity~\cite{doi:10.1126/science.aax9743,sackett2000experimental}, high-order observables~\cite{PhysRevLett.122.090503,PhysRevLett.128.150501}, or full probability distribution~\cite{doi:10.1126/science.1250147,PhysRevLett.122.173601}.
However, single-particle-resolution measurements are highly susceptible to detection noises~\cite{PhysRevLett.119.223604,PhysRevLett.115.163002}. 
To address this challenge, time-reversal interaction-based readout provides a powerful protocol to achieve high-precision Heisenberg-limited measurements~\cite{doi:10.1126/science.aaf3397,PhysRevLett.119.193601,PhysRevLett.117.013001,PhysRevLett.116.053601,PhysRevA.97.053618,PhysRevA.97.043813,PhysRevA.97.032116,colombo_time-reversal-based_2022,li2023improving,mao2023quantum,PhysRevA.110.022407}.
This requires reversing the dynamics of an interacting many-body quantum system, which is usually achieved by inverting the sign of nonlinear interaction~\cite{PhysRevLett.116.053601,PhysRevA.97.053618,PhysRevA.97.043813}.
\textit{Can one realize time-reversal dynamics to achieve Heisenberg-limited measurement without changing the sign of nonlinear interaction?}

In this Letter, we employ Floquet-engineered TAT-and-turn dynamics to efficiently create GHZ-like states and carry out the time-reversal readout without the need to invert the sign of the nonlinear interaction.
By applying a suitable transverse coupling field, an effective TAT-and-turn dynamics in an ensemble of Bose condensed two-level atoms can be realized via Floquet engineering.
Using both analytical semi-classical and quantum treatments, we find that GHZ-like states with high QFI can be generated in a much shorter time which is advantageous in mitigating the effect of decoherence.
Furthermore, the time-reversal readout can be realized by applying another Floquet-engineered TAT-and-turn dynamics without flipping the sign of nonlinear interaction, which is experimentally feasible. 
It allows us to efficiently extract the signal encoded in the prepared GHZ-like state, and thus provides a robust approach for enhancing measurement precision. 
Our work develops an efficient protocol for performing entanglement-enhanced quantum metrology via rapid generation and detection of GHZ-like states through continuous Floquet engineering.

\begin{figure}[t]
    \centering
    \includegraphics[width=1\linewidth]{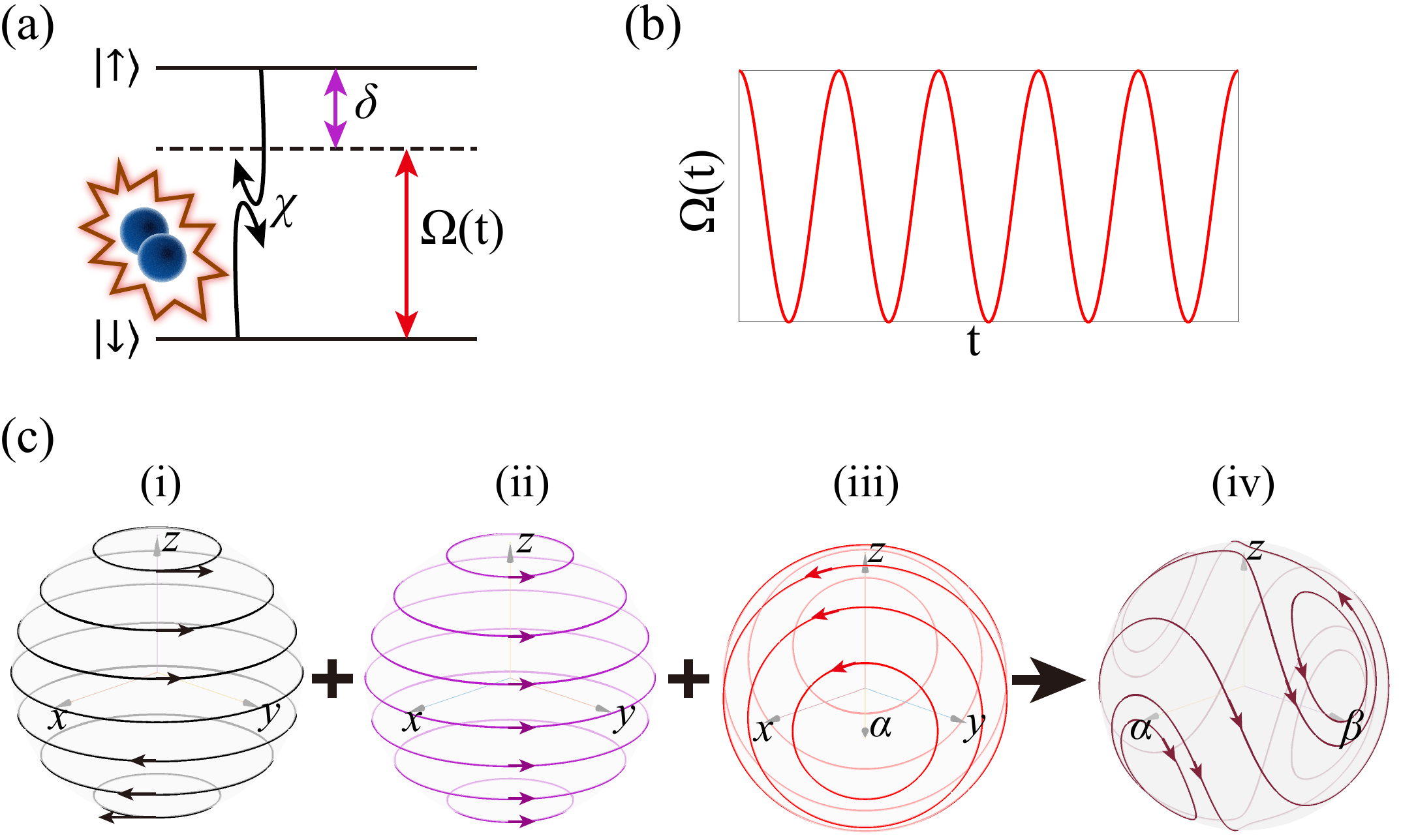}
    \caption{(a) An ensemble of two-level particles ($\chi$ denoting the nonlinear interaction between particles) coupled via an external coupling field with detuning $\delta$ and (b) periodically modulated Rabi frequency $\Omega(t)=\Omega_{0} \cos(\omega t)$. (c) Schematic diagram of Floquet-engineered TAT-and-turn. The classical phase-space trajectory for (i) one-axis-twist $\chi \hat{J}_{z}^2$, (ii) rotation induced by energy imbalance $\delta \hat{J}_{z}$, (iii) rotation induced by modulated linear coupling $\Omega_{0} \cos(\omega t)\hat{J}_{\alpha}$. The combination of these three terms results in (iv) the effective TAT-and-turn dynamics.}
    \label{fig_FE-TATNT}
\end{figure}

{\it Floquet-engineered TAT-and-turn.---}
We consider the Floquet engineering of an ensemble of Bose condensed two-level atoms coupled via Raman lasers or microwaves ~\cite{PhysRevA.59.620,PhysRevLett.95.010402,PhysRevLett.97.150402}. 
As depicted in Fig.~\ref{fig_FE-TATNT}~(a), $\chi$ represents the nonlinear interaction, $\delta$ is the detuning between the coupling field and the inter-level transition frequency (labeled by $\lvert \uparrow \rangle$ and $\lvert \downarrow \rangle$), and the Rabi frequency $\Omega(t)=\Omega_{0}\cos(\omega t)$ is periodically modulated with amplitude $\Omega_0$ and frequency $\omega$ (see Fig.~\ref{fig_FE-TATNT}~(b)). 
In the collective spin representation, the Hamiltonian can be expressed as ($\hbar = 1$ hereafter)
\begin{equation}
  \hat{H}_\mathrm{FE} = \chi \hat{J}_{z}^{2} + \delta \hat{J}_{z} + \Omega_{0} \cos(\omega t) \hat{J}_{\alpha},\label{H_FE}
\end{equation}
where $\hat{J}_{\mu} = \frac{1}{2} \sum_{k=1}^{N} \hat{\sigma}_{\mu}^{(k)} $ denote the collective spin operators with the Pauli matrices $\hat{\sigma}_{\mu}^{(k)}$ of the $k$-th particle for $\mu= x, y, z$, and $ \hat{J}_{\alpha} = \cos\alpha \hat{J}_{x}  + \sin\alpha \hat{J}_{y} $ with $\alpha$ being the phase in the $\hat{J}_{x}$-$\hat{J}_{y}$ plane.
The Hamiltonian~\eqref{H_FE} consists of three terms.
The first term, $\chi \hat{J}_{z}^{2}$, known as OAT~\cite{PhysRevA.47.5138}, generates entanglement by inducing a twisting that depends on the population imbalance.
The second term, $\delta \hat{J}_{z}$, represents the energy imbalance between $\lvert \uparrow \rangle$ and $\lvert \downarrow \rangle$, leading to a rotation around the $\hat{J}_z$ axis at a constant rate $\delta$.
The third term, $\Omega_{0} \cos(\omega t)\hat{J}_{\alpha}$, describes the Floquet-modulated coupling between $\lvert \uparrow \rangle$ and $\lvert \downarrow \rangle$, which causes a rotation around the $\hat{J}_{\alpha}$ axis with a time-dependent rate $\Omega_{0} \cos(\omega t)$.
The classical phase-space trajectories for these three terms are illustrated in Fig.~\ref{fig_FE-TATNT}~(c).
\begin{figure*}[t]
    \centering
    \includegraphics[width=1\linewidth]{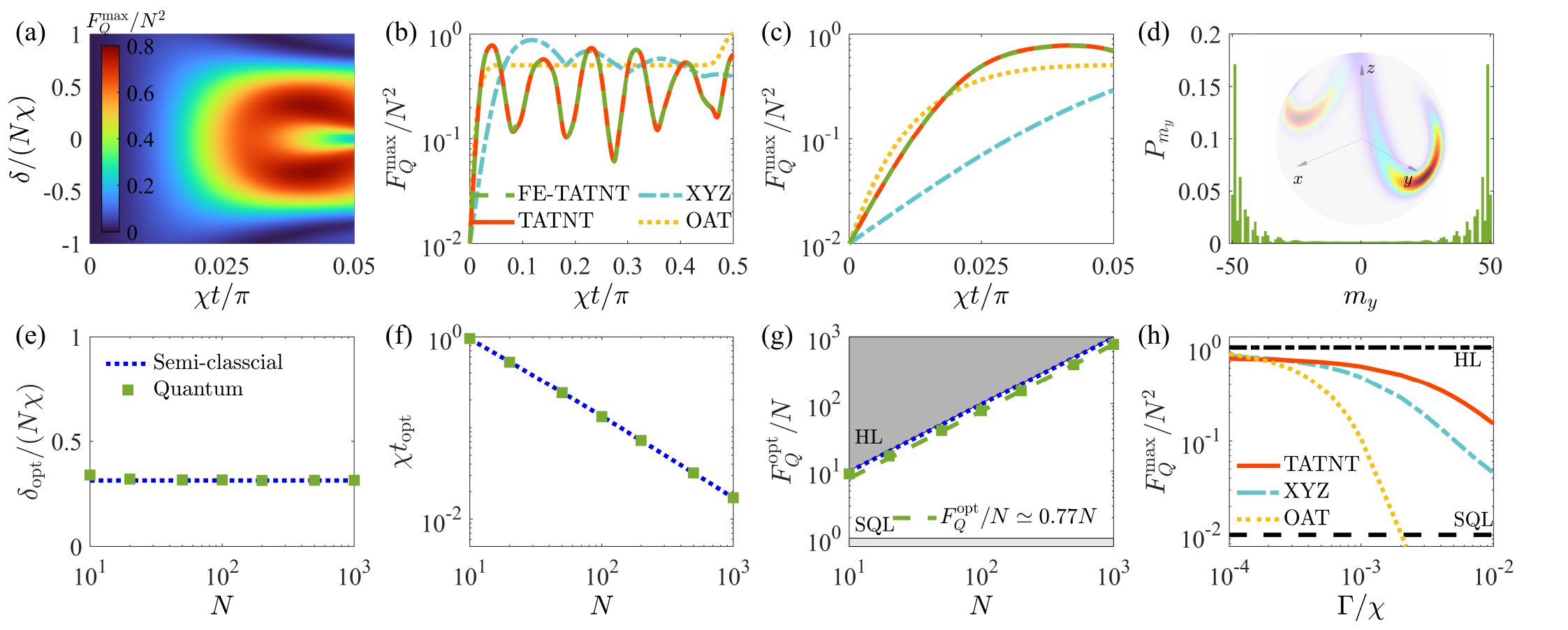} 
    \caption
    {Maximum QFI achieved via Floquet-engineered TAT-and-turn dynamics. 
    (a) The time-evolution of the maximum QFI $F_Q^\mathrm{max}$ for different $\delta$ with the particle number $N=100$, the driving frequency $\omega=2\pi\times10N\chi$ and the angle $\alpha=0$. 
    (b) The time-evolution of the maximum QFI for the Floquet-engineered TAT-and-turn with $\delta_\mathrm{opt}$. The results of OAT, XYZ~\cite{PhysRevLett.132.113402} and ideal TAT-and-turn are presented for comparison.
    (c) The enlarged area of (b) for $0\le\chi t \le 0.05\pi$.
    (d) The probability distribution $P_{m_{y}}=|\langle m_{y}|\psi\rangle|^2$ ($|m_{y}\rangle$ denoting the eigenstates of $\hat{J}_y$) and the Husimi distribution (inset) for the optimal state with largest QFI. 
    (e) The critical detuning $\delta_\mathrm{opt}$ and (f) the critical evolution time $\chi t_\mathrm{opt}$ corresponding to (g) the optimal QFI $F_Q^\mathrm{opt}$ for different particle numbers $N$, in which the semi-classical predictions are $\delta_\mathrm{opt}^\mathrm{SC}/(N\chi)\approx0.3135$, $t_\mathrm{opt}^\mathrm{SC}={3(1.9+0.55\ln{N})}/{N}$ and $(F_Q^\mathrm{opt})^\mathrm{SC}/N\approx0.97N$, respectively. 
    The fitted line for the optimal $F_Q^\mathrm{max}$ is $F_Q^\mathrm{opt}/N\simeq0.77N$. 
    (h) The optimal QFI $F_Q^\mathrm{opt}$ versus the decoherence strength $\Gamma$ in generating GHZ or GHZ-like state by OAT, XYZ, and TAT-and-turn dynamics, respectively. The black dashed (dash-dotted) line indicates the SQL (HL)}.
    \label{fig2}
\end{figure*}

By introducing a suitable energy imbalance $\delta$ and periodically modulating the coupling strength $\Omega(t)$, an effective TAT-and-turn dynamics can be achieved. 
Here, we derive the Floquet-engineered TAT-and-turn Hamiltonian from the Hamiltonian~\eqref{H_FE} using the Floquet theorem~\cite{PhysRev.138.B979, PhysRevA.68.013820}.
In the interaction picture under the unitary transformation $\hat{U} = \mathcal{\hat{T}} e^{-i \int_{0}^{t} \Omega_{0} \cos(\omega \tau) \hat{J}_{\alpha} d \tau}=e^{-i  \gamma \hat{J}_{\alpha}}$ with $\gamma = \Omega_{0} \sin (\omega t) /\omega$ and the time-ordering operator $\mathcal{\hat{T}}$,  
since $e^{i \gamma \hat{J}_{\alpha}} \hat{J}_{z} e^{-i \gamma \hat{J}_{\alpha}} = \cos \gamma \hat{J}_{z} + \sin \gamma \hat{J}_{\beta} $ with $\beta = \alpha + \pi/2$, 
the Hamiltonian~\eqref{H_FE} can be transformed into 
\begin{eqnarray}
\hat{H}_{I}&=&\hat{U}^{\dagger} \hat{H}_{\rm FE}\hat{U} -i\hat{U}^{\dagger} \partial_t \hat{U} \nonumber\\
 &=& \chi(\cos \gamma \hat{J}_{z} + \sin \gamma \hat{J}_{\beta})^2 + \delta(\cos \gamma \hat{J}_{z} + \sin \gamma \hat{J}_{\beta}). \label{H_I}
\end{eqnarray}
Using $e^{i\gamma} = \cos{\gamma} + i\sin{\gamma}$ and $e^{i z \sin \theta}= \mathcal{{J}}_{0}(|z|) + \sum_{n=1}^{\infty} [\mathcal{{J}}_{n}(z) e^{i n \theta} + (-1)^n \mathcal{{J}}_{n}(z) e^{-i n \theta} ]$ with $\mathcal{{J}}_{n}(z)$ being the Bessel function of the first kind, through performing the Fourier decomposition, the Hamiltonian~\eqref{H_I} becomes
\begin{equation}
\hat{H}_{I} = \hat{H}_{0}^{I}+\sum_{n=1}^{\infty} (\hat{H}_{n}^{I} e^{i n \omega t}+\hat{H}_{-n}^{I} e^{-i n \omega t} ),\label{H_I_Fourier}
\end{equation}
where $\hat{H}_{0}^{I} = \frac{\chi}{2} [(1+L_{0}) \hat{J}_{z}^{2} + (1 - L_{0}) \hat{J}_{\beta}^{2} ] + K_{0} \delta \hat{J}_{z}$ and $\hat{H}_{n}^{I}$ represents $n$-order Hamiltonian
with $L_{n}=\mathcal{{J}}_{n}\left(2 \Omega_{0}/\omega\right)$ and $K_{n}=\mathcal{{J}}_{n} \left(\Omega_{0}/\omega \right)$~\cite{SM}.
Finally, using the Floquet-Magnus expansion~\cite{Magnus1954, Blanes2008} up to the order of $\omega^{-1}$, the effective Hamiltonian for Eq.~\eqref{H_I_Fourier} can be written as $\hat{H}_\mathrm{F}^{\rm eff } = \hat{H}_{0}^{I} + \sum_{n=1}^{\infty}\frac{1}{n\omega} ([\hat{H}_{n}^{I},\hat{H}_{-n}^{I}]-[\hat{H}_{n}^{I},\hat{H}_{0}^{I}]+[\hat{H}_{-n}^{I},\hat{H}_{0}^{I}] )+\mathcal{O}(\omega^{-1})$.
When $\omega$ is sufficiently large ($\omega \gg N\chi$), it reduces to $\hat{H}_\mathrm{F}^{\rm eff} \simeq \hat{H}_{0}^{I}$~\cite{SM}.
For a conserved collective spin, $\hat{J}^2=\hat{J}_{\alpha}^{2}+\hat{J}_{\beta}^{2}+\hat{J}_{z}^{2}=\frac{N}{2}(\frac{N}{2}+1)$ is a constant, the effective time-independent Floquet Hamiltonian is equivalent to
\begin{equation}
\hat{H}_{\rm F} = -\frac{\chi}{2} [(1+L_{0})\hat{J}_{\alpha}^{2} + 2 L_{0} \hat{J}_{\beta}^{2} ] + K_{0} \delta \hat{J}_{z} .\label{H_F}
\end{equation}
Setting the ratio $\Omega_{0}/\omega\simeq1.6262$, we have $L_0=\mathcal{{J}}_{0}\left(2 \Omega_{0}/\omega\right)=-1/3$, thus the Hamiltonian~\eqref{H_F} becomes an effective TAT-and-turn Hamiltonian
\begin{equation}
\hat{H}_{\rm TATNT}= \chi_{\rm eff}( \hat{J}_{\beta}^{2} - \hat{J}_{\alpha}^{2}) + \delta_{\rm eff} \hat{J}_{z},\label{H_TATNT}
\end{equation}
with the effective interaction strength $\chi_\mathrm{eff}=\chi/3$ and the effective detuning $\delta_\mathrm{eff} = K_0 \delta \simeq0.4404\delta$.
Since the TAT-and-turn Hamiltonian~\eqref{H_TATNT} is derived from the Hamiltonian~\eqref{H_FE} with the Floquet theorem, we refer the original time-evolution to the Floquet-engineered TAT-and-turn (labeled as FE-TATNT) dynamics.
To evaluate the FE-TATNT dynamics, we also give the dynamics of the ideal TAT-and-turn Hamiltonian~\eqref{H_TATNT} for comparison.
Our FE-TATNT dynamics contains more general features and it can be reduced to typical TAT~\cite{PhysRevA.92.013623,PhysRevLett.129.090403}, OAT-and-turn, or OAT dynamics by varying system parameters~\cite{SM}.

{\it Fast generation of GHZ-like states using Floquet-engineered TAT-and-turn.---}
The desired entangled states are generated from from an initial state $\hat{\rho}(0)=|\psi_0\rangle\langle\psi_0|$, which is easy to prepare, according to the time-evolution $\dot{\hat{\rho}}=-i[\hat{H},\hat{\rho}]$. 
To characterize their full metrological potential, we employ the maximum QFI $F_{Q}^{\max}$, which can be derived by computing the largest eigenvalue of the QFI matrix ~\cite{PhysRevLett.131.150802,PhysRevResearch.6.033090},
\begin{equation}
\mathcal{F}\vec{n}_\mathrm{max}=F_{Q}^{\max}\vec{n}_\mathrm{max}.
\end{equation}
Here, the elements of the QFI matrix are expressed as
\begin{equation}
\mathcal{F}_{\mu\nu}=\sum_{i,j=1;\kappa_i+\kappa_j\neq0}^{\dim[\hat{\rho}]}\frac{2\mathcal{R}\left(\langle\kappa_i|[\hat{J}_\mu,\hat{\rho}]|\kappa_j\rangle\langle\kappa_j|[\hat{\rho},\hat{J}_\nu]|\kappa_i\rangle\right)}{\kappa_i+\kappa_j},
\end{equation}
with $\mu , \nu \in \{ x, y, z\}$, the spectral decomposition $\hat{\rho } =$ $\sum_i\kappa_i|\kappa_i\rangle\langle\kappa_i|$, and $\mathcal{R}$ represents the real part.
The optimal signal-encoding-direction vector is $\vec{n}_\mathrm{max}=(n_x^\mathrm{max},n_y^\mathrm{max},n_z^\mathrm{max})^\mathrm{T}$, and the responding optimal generator is $\hat{J}_{\vec{n}_\mathrm{max}}=n_x^\mathrm{max}\hat{J}_{x}+n_y^\mathrm{max}\hat{J}_{y}+n_z^\mathrm{max}\hat{J}_{z}$.
Below we use $F_Q^{\rm max}$ to analyze the full metrological potential of the states generated by the FE-TATNT dynamics~\eqref{H_FE}.

Without loss of generality, we choose $\alpha=0$ hereafter. 
It is advantageous to begin with an initial spin coherent state on one of the poles along $z$-axis ($\lvert \uparrow \rangle^{\otimes N}$ or $\lvert \downarrow \rangle^{\otimes N}$), which correspond to two saddle points $(0, 0, \pm 1)$ in the classical phase-space of TAT-and-turn when $\left\lvert \delta_{\rm eff}/\chi_{\rm eff} \right\lvert < N$, see Fig.~\ref{fig_FE-TATNT}~(c).
To identify the optimal condition for FE-TATNT, we analyze the dependence of $F_Q^{\rm max}$ on the detuning $\delta$ and the evolution time $\chi t$ in Fig.~\ref{fig2}~(a).
It clearly shows that the maximum QFI has a symmetrical distribution with respect to $\delta=0$. 
Given the particle number $N=100$, the maximum QFI reaches its optimal value $F_Q^\mathrm{opt}\simeq 0.77N^2$ when $\chi t_\mathrm{opt} \approx 0.13$ and $\delta_\mathrm{opt} \approx \pm0.3135N\chi$.
In Fig.~\ref{fig2}~(b), we show the time-evolution of $F_Q^{\rm max}$ for OAT ($\hat{H}_{\rm OAT}=\chi \hat{J}_{z}^{2}$), XYZ ($\hat{H}_{\rm XYZ}=\tilde{\alpha}\chi(\hat{J}_{x}\hat{J}_{y}\hat{J}_{z}+\hat{J}_{z}\hat{J}_{y}\hat{J}_{x})/3$ with $\tilde{\alpha}=0.4$~\cite{PhysRevLett.132.113402}), ideal TAT-and-turn ($\hat{H}_{\rm TATNT}$ with $\delta_\mathrm{eff} = K_0 \delta_\mathrm{opt}$), and FE-TATNT ($\hat{H}_{\rm FE}$ with $\Omega_{0}/\omega\simeq1.6262$ and $\delta = \delta_\mathrm{opt}$).
The result of FE-TATNT is highly consistent with the ideal TAT-and-turn, thereby confirming the validity of the effective TAT-and-turn Hamiltonian.
The FE-TATNT dynamics can attain a significantly high QFI in a very short time $\chi t \approx 0.05\pi$. 
In contrast, it requires a longer evolution time $\chi t \approx 0.5\pi$ for OAT or $\chi t \approx 0.1\pi$ for XYZ to reach the same level of QFI, see Figs.~\ref{fig2}~(b) and (c).
Furthermore, in the FE-TATNT with $\chi t_\mathrm{opt}$ and $\delta_\mathrm{opt}$, the generated state is a GHZ-like state~\cite{SM} with two main parts distributed near two poles of the $y$ axis, which is similar to a spin cat state~\cite{huang2015quantum,PhysRevA.98.012129,PhysRevA.105.062456}, see Fig.~\ref{fig2}~(d).
These results reveal that the GHZ-like state generated by FE-TATNT is much faster than that of using OAT or effective three-body interaction.

The optimal QFI, the critical detuning and the critical evolution time can be found analytically using semi-classical treatment~\cite{PhysRevA.67.013607, PRXQuantum.4.020314}.
Similarly to the spin cat state in the inset of Fig.~\ref{fig2}~(d), whose variance is approximately stretched along the $y$-axis, one can estimate the global optimal parameter $\delta_{\rm opt}^{\rm SC} = \pm (\sqrt{2}-1) N \chi / (3 K_0)\approx\pm0.3135 N \chi$ using the upper boundary of the QFI~\cite{SM}.
Due to the fact that the center of the uncertainty patch is fixed, the TAT-and-turn dynamics merely alters the distribution by stretching and squeezing. 
Thus, the relevant timescale can be estimated by the time interval between the boundary points of the uncertainty patch along the separatrix~\cite{PRXQuantum.4.020314}, which can be given as~\cite{SM}
\begin{equation}\label{timescale}
\chi t_\mathrm{opt}^\mathrm{SC} \simeq \frac{3(1.9+0.55\ln{N})}{N}.
\end{equation}
Based on the above analysis, the best achievable QFI exhibits the Heisenberg scaling, which matches the ultimate precision bound of the spin cat states~\cite{SM}.
As shown in Figs.~\ref{fig2}~(e)-(g), the above results are in accordance with the numerical results of FE-TATNT.

According to Eq.~\eqref{timescale}, the TAT-and-turn dynamics exhibits a significantly shorter timescale for large particle numbers, which is advantageous in mitigating the decoherence effect.
Here we consider the decoherence effect of spontaneous emission. 
The time-evolution can be described by~\cite{SM} $\dot{\hat{\rho}}=-i[\hat{H},\hat{\rho}]+\left(\hat{L} \hat{\rho} \hat{L}^\dagger-\frac{1}{2}\hat{L}^\dagger \hat{L}\hat{\rho}-\frac{1}{2}\hat{\rho} \hat{L}^\dagger \hat{L}\right)$ where the Lindblad operator is $\hat{L}=\sqrt{\Gamma/2}\hat{J}_{-}$ with $\hat{J}_{-}=\hat{J}_{x}-i\hat{J}_{y}$ and the spontaneous emission rate $\Gamma$~\cite{PhysRevA.4.1791,doi:10.1126/science.aar3102,PhysRevLett.132.113402}. 
As shown in Fig.~\ref{fig2}~(h), we respectively employ OAT, XYZ and TATNT dynamics to generate GHZ or GHZ-like state under decoherence, showing that the TATNT protocol is more robust against decoherence than the OAT and XYZ protocols, and demonstrating remarkable metrological usefulness in quantum metrology with noises.
Similar results for correlated dephasing are also shown in the Supplementary Material~\cite{SM}.
\begin{figure}[t]
    \centering
    \includegraphics[width=1\linewidth]{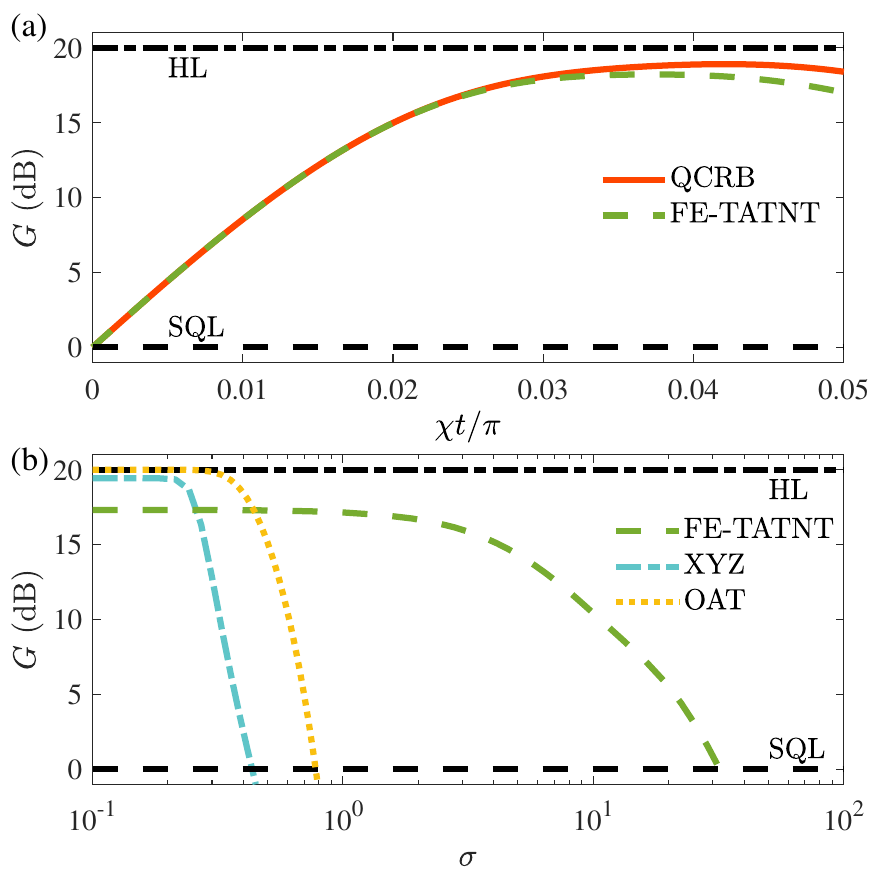} 
    \caption
    {(a) Measurement precision yielded from Floquet-engineered anti-TAT-and-turn. The metrological gain is defined as $G=20 \log _{10} \left[{(\Delta \phi)_{\mathrm{SQL}}}/{\Delta \phi}\right]$ with $(\Delta \phi)_{\mathrm{SQL}}=1/{\sqrt{N}}$ denoting the measurement precision of SQL.
    (b) Robustness against detection noise. The green dashed line is the result of population measurement $\langle\hat J_{z}\rangle_{\sigma}$ after Floquet-engineered anti-TAT-and-turn with $\chi t=0.12$. The blue dash-dotted and yellow dotted lines are results of parity measurement $\langle\hat{\prod}\rangle_{\sigma}$ for GHZ-like and GHZ states generated by XYZ and OAT dynamics, respectively.     
    Here, the black dashed (dash-dotted) line indicates the SQL (HL). The driving frequency is set to $\omega = 2\pi \times 100N\chi$, and $\alpha = \pi/2$. These parameters, along with a particle number of $N = 100$ and an estimated phase of $\phi = 1/1000$, are selected for the simulation of FE-TATNT.}
    \label{fig3}
\end{figure}

{\it Interaction-based readout without flipping the sign of nonlinear interaction.---}
It is crucial to approach the ultimate precision bound (i.e. the QCRB) set by the QFI. 
For a tiny estimated phase, time-reversal readout may approach the QCRB without single-particle resolved detection~\cite{PhysRevLett.115.163002,doi:10.1126/science.aaf3397,PhysRevLett.119.193601,PhysRevLett.117.013001,PhysRevLett.119.223604,PhysRevLett.116.053601,PhysRevA.97.053618,PhysRevA.97.043813,PhysRevA.97.032116,doi:10.1126/science.aaw2884,doi:10.1126/science.abi5226,colombo_time-reversal-based_2022,li2023improving,mao2023quantum}.
Here, we find that the optimal measurements can be achieved via anti-TAT-and-turn, regarded as the time-reversal process of TAT-and-turn. 
More importantly, the anti-TAT-and-turn can be realized by Floquet engineering without the need of flipping the sign of nonlinear interaction $\chi$. 

With the generated GHZ-like state, one can encode the estimated phase $\phi$ along the best sensing direction by $\hat{R}_{\vec{n}_{\rm max}}(\phi) = e^{-i \phi \hat{J}_{\vec{n}_{\rm max}}}$. 
Here, $\phi$ is accumulated during free evolution and depends on the physical quantity to be measured. 
The physical quantity can be the frequency detuning of a reference laser driving the atoms in atomic clocks or the magnetic field responsible for Zeeman splitting of the atomic levels in magnetometers.
Based on the Hamiltonian~\eqref{H_FE}, by adjusting the detuning $\delta \mapsto -\delta$ and the initial phase of external coupling field $\alpha=0 \mapsto \alpha=\pi/2$, one can obtain an effective anti-TAT-and-turn Hamiltonian $\hat{H}_{\rm {TATNT}}^{\alpha=\pi/2,-\delta}=-\hat{H}_{\rm{TATNT}}^{\alpha=0,\delta}$.
Ideally, the state before measurement can be written as
\begin{eqnarray}\label{final_state}
    \lvert\psi_{f}\rangle &=& \hat{U}_2 \hat{R}_{\vec{n}}(\phi) \hat{U}_1\lvert\psi_{i}\rangle \nonumber\\
    &=& e^{-i \int \hat{H}_{\rm FE}^{\alpha=\pi/2,-\delta} t} e^{-i \phi \hat{J}_{\vec{n}_{\rm max}}} e^{-i \int \hat{H}_{\rm FE}^{\alpha=0,\delta} t} \lvert\psi_{i}\rangle  \nonumber\\
    &\approx& e^{-i \hat{H}_{\rm {TATNT}}^{\alpha=\pi/2, -\delta} t} e^{-i \phi \hat{J}_{\vec{n}_{\rm max}}} e^{-i \hat{H}_{\rm TATNT}^{\alpha=0,\delta} t} \lvert\psi_{i}\rangle, 
\end{eqnarray}
where $\lvert\psi_{i}\rangle=\lvert \uparrow \rangle^{\otimes N}$, $\hat{H}_{\rm TATNT}^{\alpha=0,\delta}=\chi_{\rm eff}( \hat{J}_{y}^{2} - \hat{J}_{x}^{2}) + \delta_{\rm eff} \hat{J}_{z}$ and $\hat{H}_{\rm {TATNT}}^{\alpha=\pi/2,-\delta}=-\chi_{\rm eff}( \hat{J}_{y}^{2} - \hat{J}_{x}^{2}) - \delta_{\rm eff} \hat{J}_{z}$.
Furthermore, to find out the optimal measurement, one can optimize the sensitivity over all possible directions through singular value decomposition 
\begin{equation}\label{opt_measure}
    \mathop{\max}_{\vec{m}}[\Delta\phi(\vec{n}_{\rm max}, \vec{m})]^{-2}  = \vec{n}_{\rm max}^T \mathcal{K} \vec{n}_{\rm max},
\end{equation}
where $\mathcal{K} = \mathcal{M} \mathcal{Q}^{-1} \mathcal{M}^T$ with two $3 \times 3$ matrices $\mathcal{M}$ and $\mathcal{Q}$, whose elements are given as $\mathcal{M}_{\mu\nu} = i\langle [\hat{U}^\dagger_1 \hat{J}_{\mu} \hat{U}_1 , \hat{\tilde{J}}_{\nu}]\rangle_i$ and $\mathcal{Q}_{\mu\nu} = \langle \{\hat{\tilde{J}}_{\mu}, \hat{\tilde{J}}_{\nu}\} \rangle_i/2 - \langle\hat{\tilde{J}}_{\mu}\rangle_i \langle\hat{\tilde{J}}_{\nu}\rangle_i$ for $\mu, \nu= x, y, z$ with $\hat{\tilde{J}}_{\mu(\nu)} = \hat{U}^\dagger_1 \hat{R}_{\vec{n}_{\rm max}}^\dagger(\phi) \hat{U}^\dagger_2 \hat{J}_{\mu(\nu)} \hat{U}_2 \hat{R}_{\vec{n}_{\rm max}}(\phi) \hat{U}_1$~\cite{Schulte2020ramsey, PhysRevA.107.052613}.
The optimal measurement direction can then be given as $\vec{m}_{\rm max} = \mathcal{N}_c \mathcal{Q}^{-1} \mathcal{M}^T \vec{n}$ with the normalization constant $\mathcal{N}_c$ determined by the Cauchy-Schwarz inequality~\cite{PhysRevLett.122.090503}.
Therefore, one can easily rotate the final state and measure the half-population difference $\hat J_z$ to infer $\phi$.
In particular, the above calculations are valid for an arbitrary phase $|\phi| \ge 0$, which generalizes the result only for the zero phase $\phi=0$~\cite{SM}.
As shown in Fig.~\ref{fig3}~(a), the nonlinear readout without flipping the sign of nonlinear interaction for FE-TATNT achieves excellent measurement precision approaching the corresponding QCRB.

Furthermore, an important capability of the Floquet-engineered anti-TAT-and-turn readout is its robustness against detection noise.
Here, we assume the detector possessing Gaussian noise and the probability distribution under detection noise is described by $P_{m}(\phi|\sigma)=\sum_{n=-N/2}^{N/2} C_{n} e^{-(m-n)^{2}/2 \sigma^{2}} P_{n}(\phi)$ with the normalization $C_n=1/\sum_{m=-N/2}^{N/2} \mathrm{e}^{-(m-n)^{2}/2 \sigma^{2}}$ and the ideal probability distribution $P_{n}(\phi)=\vert\langle n\lvert\psi_{f}(\phi)\rangle\vert^2$~\cite{PhysRevLett.119.193601}.
Under detection noise, the outcome is expressed as $\langle{\hat{J}}_{z}\rangle_{\sigma}=\sum_{m=-N/2}^{N/2} mP_{m}(\phi|\sigma)$ for population measurement or $\langle\Pi\rangle_{\sigma}=\sum_{m=-N/2}^{N/2} (-1)^{m}P_{m}(\phi|\sigma)$ for parity measurement~\cite{PhysRevA.97.053618}.
As shown in Fig.~\ref{fig3}~(b), for $\sigma \ge 1$, while the parity measurement becomes invalid, our protocol can still achieve a metrological gain beyond SQL, showing that the robustness of our protocol is much better than that of using the parity measurement.
Our Floquet-engineered anti-TAT-and-turn can be achieved by adjusting the periodic coupling without changing the interaction, which is feasible for realistic entanglement-enhanced sensing.

{\it Conclusions.---}
We use continuous Floquet engineering with periodically modulated Rabi frequency under appropriate detuning to achieve an effective TAT-and-turn dynamics.
With semi-classical and quantum approaches, we analytically find the preparation timescale at large particle number is significantly shorter compared to the OAT, and even faster than using artificial three-body interactions, which is advantageous in mitigating decoherence.
Moreover, applying another Floquet-engineered TAT-and-turn dynamics, one can realize the time-reversal readout without flipping the sign of the nonlinear interaction. 
Our work provides an alternative way for fast generation and detection of large-particle-number GHZ-like states, which can be created in various systems~\cite{hosten2016measurement,PRXQuantum.3.020308,doi:10.1126/science.1208798,doi:10.1126/science.aag1106,franke2023quantum,doi:10.1126/science.aad9958,eckner2023realizing,panda2024measuring,doi:10.1126/sciadv.abg9204,PhysRevLett.132.240803} and has promising applications in quantum metrology and quantum information science.

\begin{acknowledgments}
The authors thank Dr. Wenjie Liu for helpful discussions. This work is supported by the National Natural Science Foundation of China (Grants No. 12025509 and No. 92476201), the National Key Research and Development Program of China (Grant No. 2022YFA1404104), and the Guangdong Provincial Quantum Science Strategic Initiative (GDZX2305006 and GDZX2405002).
\end{acknowledgments}

\nocite{}

  \begin{widetext}
    \begin{center}
    \textbf{\Large Supplemental Material}
    \end{center}
    \section{I. Derivation of Floquet-engineered TAT-and-turn Hamiltonian}
    By periodically modulating an external field that couples to an ensemble of identical Bose-condensed two-level atoms, the Floquet-engineered Hamiltonian can be expressed as:
    \begin{equation}
      \hat{H}_{\rm FE} = \chi \hat{J}_{z}^{2} + \delta \hat{J}_{z} + \Omega(t) \hat{J}_{\alpha}, \label{H_FE}
    \end{equation}
    where $\Omega(t) = \Omega_0 \cos(\omega t)$, and $\hat{J}_{\alpha} = \hat{J}_{x} \cos\alpha + \hat{J}_{y} \sin\alpha$. 
    To facilitate the analytical calculation, we first transform from the Schrödinger picture to the interaction picture, where $\hat{H}_{\rm FE}$ is decomposed into a time-independent term, $\hat{H}_{0} = \chi \hat{J}_{z}^{2} + \delta \hat{J}_{z}$, and a time-dependent term, $\hat{V}(t) = \Omega_{0} \cos(\omega t) \hat{J}_{\alpha} $.
    Applying the Baker-Campbell-Hausdorff formula, we obtain $e^{i \gamma \hat{J}_{\alpha}} \hat{J}_{z} e^{-i \gamma \hat{J}_{\alpha}} = \cos \gamma \hat{J}_{z} + \sin \gamma \hat{J}_{\beta} $ where $\beta = \alpha + \pi/2$. 
    In the interaction picture, we derive
    \begin{equation}
    \begin{aligned}
    \hat{H}_{I} = & \hat{U}^{\dagger} \hat{H}_{\rm FE}\hat{U} -i\hat{U}^{\dagger} \partial_t \hat{U} \\
    =&\hat{U}^{\dagger} \hat{H}_{0} \hat{U} \\
    =& e^{ i \int_{0}^{t} \Omega_{0} \cos(\omega \tau) \hat{J}_{\alpha} d \tau} \chi \hat{J}_{z}^{2} e^{-i \int_{0}^{t} \Omega_{0} \cos(\omega \tau) \hat{J}_{\alpha} d \tau} +e^{i \int_{0}^{t} \Omega_{0} \cos(\omega \tau) \hat{J}_{\alpha} d \tau} \delta \hat{J}_{z} e^{-i \int_{0}^{t} \Omega_{0} \cos(\omega \tau) \hat{J}_{\alpha} d \tau} \\
    = & \chi(\hat{J}_{z} \cos \gamma + \hat{J}_{\beta} \sin \gamma )^2 + \delta(\hat{J}_{z} \cos \gamma + \hat{J}_{\beta} \sin \gamma ),
    \end{aligned}
    \end{equation}
    where $\hat{U} = \mathcal{\hat{T}} e^{-i \int_{0}^{t} \Omega_{0} \cos(\omega \tau) \hat{J}_{\alpha} d \tau}=e^{-i  \gamma \hat{J}_{\alpha}}$, with $\gamma = \Omega_{0} \sin (\omega t) /\omega$ and $\mathcal{\hat{T}}$ is the time-ordering operator. 

    Using the Euler formula $e^{i\gamma} = \cos{\gamma} + i\sin{\gamma}$ and the Jacobi-Anger expansion variant $e^{i z \sin \theta}= \mathcal{{J}}_{0}(|z|) + \sum_{n=1}^{\infty}\left[ \mathcal{{J}}_{n}(z) e^{i n \theta} + (-1)^n \mathcal{{J}}_{n}(z) e^{-i n \theta}\right]$ , where  $\mathcal{{J}}_{n}(z)$  is the $n$-th Bessel function of the first kind, we perform the Fourier decomposition and the Hamiltonian can be expressed as
    \begin{equation}
    \begin{aligned}
    \hat{H}_{I} = \hat{H}_{0}^{I}+\sum_{n=1}^{\infty} (\hat{H}_{n}^{I} e^{i n \omega t}+\hat{H}_{-n}^{I} e^{-i n \omega t} )
    \label{H_I_Fourier}
    \end{aligned}
    \end{equation}
    where the $n$-order terms in the Hamiltonian are given by
    \begin{equation}
    \begin{aligned}
    \hat{H}_{0}^{I} =& \frac{\chi}{2} [(1+L_{0}) \hat{J}_{z}^{2} + (1 - L_{0}) \hat{J}_{\beta}^{2} ] + K_{0} \delta \hat{J}_{z}, \\
    \hat{H}_{n}^{I}=&\frac{\chi}{4}L_{n}[(-1)^n \hat{J}_{1}^{2} + \hat{J}_{2}^{2}] + \frac{\delta}{2}K_{n}[(-1)^n \hat{J}_{1} + \hat{J}_{2}],\\
    \hat{H}_{-n}^{I}=&\frac{\chi}{4}L_{n}[ \hat{J}_{1}^{2} + (-1)^n \hat{J}_{2}^{2}] + \frac{\delta}{2}K_{n}[\hat{J}_{1} + (-1)^n \hat{J}_{2}],
    \end{aligned}
    \label{H_Fourier}
    \end{equation}
    with $\hat{J}_{1,2}=\hat{J}_{z} \pm i \hat{J}_{\beta}$, $L_{n}=\mathcal{{J}}_{n}\left(2 \Omega_{0}/\omega\right)$ , $K_{n}=\mathcal{{J}}_{n} \left(\Omega_{0}/\omega \right)$ and $n = 1,2,3 \cdots$.
    When $n$ is odd (denoted by $p$), we have $H_{-p}^{I}=-H_{p}^{I}$, so the commutation relations between Hamiltonians are
    \begin{equation}
    \begin{aligned}
        &[\hat{H}_{p}^{I},\hat{H}_{-p}^{I}]=0,\\
        &[\hat{H}_{p}^{I},\hat{H}_{0}^{I}]=a_{p}\chi^2\{\hat{J}_{z},\{\hat{J}_{z},\hat{J}_{\alpha}\}\} + b_{p}\chi^2\{\hat{J}_{\beta},\{\hat{J}_{\beta},\hat{J}_{\alpha}\}\}+c_{p}\chi\delta\{\hat{J}_{\alpha},\hat{J}_{z}\} + d_{p}\delta^2\hat{J}_{\alpha},\\
        &[\hat{H}_{-p}^{I},\hat{H}_{0}^{I}]=-[\hat{H}_{p}^{I,1},\hat{H}_{0}^{I}].
    \end{aligned}
    \end{equation}
    with 
    \begin{equation}
    \begin{aligned}
        a_{p}= \frac{1}{4} L_{p} (L_{0}+1),~ b_{p}= \frac{1}{4} L_{p} (L_{0}-1),~
        c_{p}=\frac{1}{2} (L_{p} K_{0} + L_{0} K_{p} + K_{p}),~ d_{p}= K_{0} K_{p},
    \end{aligned}
    \end{equation}
    where $[\hat{P},\hat{Q}]=\hat{P}\hat{Q}-\hat{Q}\hat{P}$ and $\{\hat{P},\hat{Q}\}=\hat{P}\hat{Q}+\hat{Q}\hat{P}$ are commutation and anti-commutation, respectively.
    For even $n$ (denoted by $q$), we have $H_{-q}^{I}=H_{q}^{I}$, and the commutation relations become
    \begin{equation}
    \begin{aligned}
    [\hat{H}_{q}^{I},\hat{H}_{-q}^{I}] &= 0,\\ 
    [\hat{H}_{-q}^{I},\hat{H}_{0}^{I}] &= [\hat{H}_{q}^{I},\hat{H}_{0}^{I}].
    \end{aligned}
    \end{equation}
    
    According to the Floquet-Magnus expansion~\cite{Magnus1954, Blanes2008}, the  effective Floquet Hamiltonian becomes
    \begin{equation}
    \begin{aligned}
    \hat{H}_\mathrm{F}^{\rm eff} &= \hat{H}_{0}^{I} + \sum_{n=1}^{\infty}(\frac{[\hat{H}_{n}^{I},\hat{H}_{-n}^{I}]}{n\omega}-\frac{[\hat{H}_{n}^{I},\hat{H}_{0}^{I}]}{n\omega}+\frac{[\hat{H}_{-n}^{I},\hat{H}_{0}^{I}]}{n\omega}) + \mathcal{O}(\omega^{-1}) \\
    &=\hat{H}_{0}^{I}-\sum_{p=1}^{\infty}\frac{2}{p\omega}[\hat{H}_{p}^{I},\hat{H}_{0}^{I}] + \mathcal{O}(\omega^{-1}) \\
    &=\hat{H}_{0}^{I}-\frac{2\chi^2}{\omega}\{\hat{J}_{z},\{\hat{J}_{z},\hat{J}_{\alpha}\}\}\sum_{p=1}^{\infty}\frac{a_p}{p} -\frac{2\chi^2}{\omega}\{\hat{J}_{\beta},\{\hat{J}_{\beta},\hat{J}_{\alpha}\}\}\sum_{p=1}^{\infty}\frac{b_p}{p} -\frac{2\chi\delta}{\omega}\{\hat{J}_{\alpha},\hat{J}_{z}\}\sum_{p=1}^{\infty}\frac{c_p}{p}-\frac{2\delta^2}{\omega}\hat{J}_{\alpha}\sum_{p=1}^{\infty}\frac{d_p}{p} + \mathcal{O}(\omega^{-1}).
    \end{aligned}
    \label{H_Floquet1}
    \end{equation}
    It indicates that the $q$(even)-order terms do not contribute to the effective Floquet Hamiltonian.
    In order to retain only $\hat{H}_{0}^{I}$ in the Hamiltonian~\eqref{H_Floquet1}, corresponding to drop the time-dependent terms in the Hamiltonian~\eqref{H_I_Fourier}, we require $\omega$ to be sufficiently large $(\omega \gg N\chi)$. Under this condition, the Hamiltonian simplifies to
    \begin{equation}
    \hat{H}_{\rm F}^{\mathrm{eff}} \simeq \hat{H}_{0}^{I}= \frac{\chi}{2} [(1+L_{0}) \hat{J}_{z}^{2} + (1 - L_{0}) \hat{J}_{\beta}^{2} ] + K_{0} \delta \hat{J}_{z},
    \end{equation}
    which reveals that the external driving field generates twisting along both the $z$ and $\beta$ directions.
    Assuming that the $\hat{J}^2 = \hat{J}_{\alpha}^{2} + \hat{J}_{\beta}^{2} + \hat{J}_{z}^{2}$ is conserved during the dynamical evolution, we derive a Hamiltonian that combines features of both OAT and TAT Hamiltonians by incorporating an additional term $-\frac{\chi}{2}(1+L_{0}) \hat{J}^{2}$. The resulting Hamiltonian can be expressed as:
    \begin{equation}
    \begin{aligned}
    \hat{H}_{\rm F} &= -\frac{\chi}{2} [(1+3L_{0})\hat{J}_{\alpha}^{2} + 2 L_{0} ( \hat{J}_{\beta}^{2} - \hat{J}_{\alpha}^{2}) ] + K_{0} \delta \hat{J}_{z} \\
    &= -\frac{\chi}{2} [(1+L_{0})\hat{J}_{\alpha}^{2} + 2 L_{0} \hat{J}_{\beta}^{2} ] + K_{0} \delta \hat{J}_{z}.
    \end{aligned}
    \end{equation}
    When $L_{0}=-1/3$, corresponding to ${\Omega_{0}}/{\omega}\simeq1.6262$, the Hamiltonian becomes an effective TAT-and-turn Hamiltonian:
    \begin{equation}
    \hat{H}_{\rm TATNT}= \chi_{\rm eff}( \hat{J}_{\beta}^{2} - \hat{J}_{\alpha}^{2}) + \delta_{\rm eff} \hat{J}_{z},
    \label{TATNT Hamiltonian}
    \end{equation}
    where the effective nonlinear interaction strength is $\chi_\mathrm{eff}=\chi/3$ and the effective detuning is $\delta_\mathrm{eff}=\delta{K}_{0}\simeq0.4404\delta$.
    Furthermore, by adjusting the initial phase of external coupling field and flipping the detuning ($\alpha \mapsto \alpha+\pi/2$ and $\delta \mapsto -\delta$), one can obtain the anti-TAT-and-turn Hamiltonian 
    \begin{equation}
    \hat{H}_{\rm TATNT}^{\rm anti} = \chi_{\rm eff}( \hat{J}_{\alpha}^{2} - \hat{J}_{\beta}^{2}) - \delta_{\rm eff} \hat{J}_{z},\label{anti-H_TATNT}
    \end{equation}
    which can be applied for time-reversal interaction-based readout without flipping the sign of nonlinear interaction.

    \section{II. Semi-classical treatment of TAT-and-turn dynamics}
    We use the semi-classical treatment, as introduced in Ref.~\cite{PRXQuantum.4.020314}, to analyze the metrological properties of TAT-and-turn dynamics.
    Utilizing the commutation relations $[\hat{J}_{\alpha}, \hat{J}_{\beta}] = i \epsilon_{\alpha\beta z} \hat{J}_{z}$, 
    we can describe the time-evolution of Hamiltonian~\eqref{TATNT Hamiltonian} using the Heisenberg equations of motion, $d\hat{J}_\kappa/dt=i[\hat{H},\hat{J}_\kappa]$, where $\kappa=\alpha,\beta,z$ are the components of the collective spin. The Heisenberg equations of motion are given by
    \begin{equation}
    \begin{aligned}
    \frac{d\hat{J}_\alpha}{dt}&=i[\hat{H}_\mathrm{TATNT},\hat{J}_\alpha]=\chi_{\mathrm{eff}}(\hat{J}_{\beta}\hat{J}_z+\hat{J}_z\hat{J}_{\beta})-\delta_\mathrm{eff}\hat{J}_\beta,\\ 
    \frac{d\hat{J}_\beta}{dt}&=i[\hat{H}_\mathrm{TATNT},\hat{J}_\beta]=\chi_{\mathrm{eff}}(\hat{J}_{\alpha}\hat{J}_z+\hat{J}_z\hat{J}_{\alpha})+\delta_\mathrm{eff}\hat{J}_\alpha,\\ 
    \frac{d\hat{J}_z}{dt}&=i[\hat{H}_\mathrm{TATNT},\hat{J}_z]=-2\chi_{\mathrm{eff}}(\hat{J}_{\alpha}\hat{J}_{\beta}+\hat{J}_{\beta}\hat{J}_{\alpha}).
    \end{aligned}
    \label{SCH}
    \end{equation}
    In the thermodynamic limit, $J=N/2\to\infty$, above Eq.(\ref{SCH}) leads to the phase-space flow of the classical variables $\mathbf{R}=(A, B, Z)=$ $(\langle\hat{{J}}_\alpha\rangle, \langle\hat{{J}}_\beta\rangle, \langle\hat{{J}}_z\rangle)/J$ with $A^2 + B^2 + Z^2 = 1$.
    After neglecting correlations $\langle\hat{P}\hat{Q}\rangle=\langle\hat{P}\rangle\langle\hat{Q}\rangle$, the equations are given by
    \begin{equation}
    \begin{aligned}
    \frac{dA}{dt}&=N\chi_\mathrm{eff}BZ-\delta_{\mathrm{eff}}B,\\ 
    \frac{dB}{dt}&=N\chi_\mathrm{eff}AZ+\delta_{\mathrm{eff}}A,\\ 
    \frac{dZ}{dt}&=-2N\chi_\mathrm{eff}AB. 
    \label{CE}
    \end{aligned}
    \end{equation}
    Likewise, the classical phase-space trajectories of nonlinear OAT or linear modulation can be obtained using the above method and are plotted in Fig.1~(c) of the main text.

    The direction of the initial spin coherent state in the non-adiabatic evolution of the Hamiltonian~\eqref{TATNT Hamiltonian} is determined by the fixed points of a phase-space flow with a trivial evolution. By solving $d\mathbf{R} / dt= 0$, the phase-space flow in Eq.~\eqref{CE} has six different fixed points depending on the system parameters (we consider the conditions $\chi \neq 0$, and $|\delta_{\mathrm{eff}} /( N \chi_{\mathrm{eff}})|< 1$), which are given by
    \begin{equation}
    \begin{aligned}
    (A,B,Z)&=(0,0,\pm1),\\ 
    (A,B,Z)&=\left(0,\pm\sqrt{1-\left(\frac{\delta_{\mathrm{eff}}}{N\chi_{\mathrm{eff}}}\right)^2},\frac{\delta_{\mathrm{eff}}}{N\chi_{\mathrm{eff}}}\right),\\ 
    (A,B,Z)&=\left(\pm\sqrt{1-\left(\frac{\delta_{\mathrm{eff}}}{N\chi_{\mathrm{eff}}}\right)^2},0,-\frac{\delta_{\mathrm{eff}}}{N\chi_{\mathrm{eff}}}\right).
    \end{aligned}
    \end{equation}

    For the fixed points, the stability analysis is carried out by diagonalizing of the Jacobi matrix, ${M}[\mathbf{R}]=(\partial/\partial\mathbf{R})(d\mathbf{R}/dt)$. In Eq.~\eqref{CE}, the matrix of the phase-space flow is given by
    \begin{equation}
    {M}[\mathbf{R}] = \begin{pmatrix}
    0 && N\chi_{\mathrm{eff}}Z-\delta_{\mathrm{eff}} && N\chi_{\mathrm{eff}}B \\ N\chi_{\mathrm{eff}}Z+\delta_{\mathrm{eff}} && 0 && N\chi_{\mathrm{eff}}A \\ -2N\chi_{\mathrm{eff}}B && -2N\chi_{\mathrm{eff}}A && 0
    \end{pmatrix}.
    \end{equation}
    The analysis shows that two fixed points $(A,B,Z)=(0,0,\pm1)$ which always point along the $\pm z$-axis are saddle points, while the other four points are stable. The Jacobi matrix of the saddle points are
    \begin{equation}
    {M}[\mathbf{R}]=\begin{pmatrix}0&&\pm N\chi_{\mathrm{eff}}-\delta_{\mathrm{eff}}&&0\\\pm N\chi_{\mathrm{eff}}+\delta_{\mathrm{eff}}&&0&&0\\0&&0&&0\end{pmatrix}.
    \label{JM_FP}
    \end{equation}

    In general, the separatrix branches emerge from this saddle point, which yield the classical evolution trajectory of the Hamiltonian. By evaluating the eigenvalues of the Jacobi matrix in Eq.~\eqref{JM_FP} at this saddle, one obtains its local Lyapunov exponent:
    \begin{equation}
    \Lambda_{\mathrm{sd}}^{\mathrm{TATNT}}=N\chi_{\mathrm{eff}}\sqrt{1-\biggl(\frac{\delta_{\mathrm{eff}}}{N\chi_{\mathrm{eff}}}\biggr)^2}.
    \label{LE_TATNT}
    \end{equation}
    It dictates the exponential rate at which points move away from the saddle point, i.e., the optimal entanglement generation parameters at the initial moment. 
    This rate reaches its maximum when $\delta_{\mathrm{eff}}=0$, which corresponds to the TAT Hamiltonian and the corresponding Lyapunov exponent is given by $\Lambda_{\mathrm{sd}}^{\mathrm{TAT}} = N\chi_{\mathrm{eff}}$. 

    Besides, we can obtain the energy density of the TAT-and-turn Hamiltonian by computing $\langle\hat{H}_{\mathrm{TATNT}}\rangle/J$ in the thermodynamic limit. After neglecting correlations as before, and in terms of the classical variables $\mathbf{R} = (A, B, Z)$, the energy density in classical phase space becomes
    \begin{equation}
    E(A,B,Z;\delta_{\mathrm{eff}},\chi_{\mathrm{eff}})=\frac{\langle\hat{H}_{\mathrm{TATNT}}\rangle}{J}=\delta_{\mathrm{eff}} Z-\frac{N\chi_{\mathrm{eff}}}{2}(A^{2}-B^{2}).
    \end{equation}
    Since the conservation of energy ensures that points on the separatrix have the same energy, the separatrix equation of the two saddles can be constructed as $E(A,B,Z;\delta_{\mathrm{eff}},\chi_{\mathrm{eff}}) = E(0,0,\pm 1;\delta_{\mathrm{eff}},\chi_{\mathrm{eff}})= \pm \delta_{\mathrm{eff}}$. 
    Due to the constraint $A^{2}+B^{2}+Z^{2}=1$, the above formula can be rewritten as
    \begin{equation}
    E(A,B,Z;\delta_{\mathrm{eff}},\chi_{\mathrm{eff}})=\delta_{\mathrm{eff}} Z-\frac{N\chi_{\mathrm{eff}}}{2}(2A^{2}+Z^{2}-1)=\pm\delta_{\mathrm{eff}},
    \end{equation}
    or
    \begin{equation}
    E(A,B,Z;\delta_{\mathrm{eff}},\chi_{\mathrm{eff}}) = \delta_{\mathrm{eff}} Z-\frac{N\chi_{\mathrm{eff}}}{2}(1-Z^{2}-2B^{2}) = \pm\delta_{\mathrm{eff}}.
    \end{equation}
    The relationships between classical space components are obtained as
    \begin{equation}
    A^{2} = \frac{1}{2}(1 \mp Z)(1 \pm Z \mp \frac{2\delta_{\mathrm{eff}}}{N\chi_{\mathrm{eff}}}),
    \end{equation}
    and 
    \begin{equation}
    B^{2} = \frac{1}{2}(1 \mp Z)(1 \pm Z \pm \frac{2\delta_{\mathrm{eff}}}{N\chi_{\mathrm{eff}}}).
    \end{equation}

    We can further estimate the globally optimal parameter $\delta_{\rm opt}$ and the explicit expression for the time evolution to reach maximum quantum Fisher information (QFI) using the semi-classical description. We find that the TAT-and-turn Hamiltonian can generate GHZ-like states, whose variance is stretched along the $\alpha$-axis or $\beta$-axis on the $x-y$ plane. For calculation clarity, we start from the initial position $(0, 0, 1)$. 
    When $\delta_{\mathrm{eff}} > 0$ (with the default assumption that $\chi > 0$), the QFI is given by
    \begin{equation}
    F_Q=4\left(\Delta \hat{J}_\beta\right)^2 = N^2B^2 = \frac{N^2}{2}(1-Z)(1+Z+\frac{2\delta_{\mathrm{eff}}}{N\chi_{\mathrm{eff}}}),\label{qfi_c}
    \end{equation}
    where we take $\hat{J}_\beta$ as the generator.
    When $Z=-\frac{\delta_{\mathrm{eff}}}{N\chi_{\mathrm{eff}}}$, the above Eq.~\eqref{qfi_c} reaches its maximum as a function of $Z$, and it simplifies to
    \begin{equation}
    F_Q=\frac{N^2}{2}(1+\frac{\delta_{\mathrm{eff}}}{N\chi_{\mathrm{eff}}})^2 \leq (F_Q)_\mathrm{max} \leq N^2.
    \end{equation}
    The maximum QFI $(F_Q)_\mathrm{max}$ occurs when 
    \begin{equation}
    {\delta_{\rm opt}}/({N\chi_{\mathrm{eff}}}) = \sqrt{2}-1 \approx 0.414. \label{detuning_c}
    \end{equation}
    For $\delta_{\mathrm{eff}} < 0$, the QFI is expressed as $F_Q=4\left(\Delta \hat{J}_\alpha\right)^2=N^2A^2\leq N^2$ with the generator being $\hat{J}_\alpha$.
    In this case, the optimal detuning is ${\delta_{\rm opt}}/({N\chi_{\mathrm{eff}}}) =1-\sqrt{2} \approx -0.414$.

    We consider $\delta_{\mathrm{eff}} > 0$ (with the analogous calculation for $\delta_{\mathrm{eff}} < 0$ provided below), within the time range $0 \leq t \leq t_{\rm opt}$ of reaching the maximum QFI. We find that $AB=\frac{1}{2}(1-Z)\sqrt{(1+Z)^2-(\frac{2\delta_{\mathrm{eff}}}{N\chi_{\mathrm{eff}}})^2} > 0$, leading to a differential equation that depends solely on the variable $Z$, which is given by
    \begin{equation}
    \frac{dZ}{dt}=-2N\chi_{\mathrm{eff}}AB=-N\chi_{\mathrm{eff}}(1-Z)\sqrt{(1+Z)^2-(\frac{2\delta_{\mathrm{eff}}}{N\chi_{\mathrm{eff}}})^2},
    \end{equation}
    and the desired timescale is given by the integration
    \begin{equation}
    \begin{aligned}
        N\chi_{\mathrm{eff}}t_{\rm opt}=-\int_{Z(0)}^{Z(t_{\rm opt})} \frac{dZ}{(1-Z)\sqrt{(1+Z)^2-(\frac{2\delta_{\mathrm{eff}}}{N\chi_{\mathrm{eff}}})^2}} &=-\frac{f(Z)}{\sqrt{1-(\frac{\delta_{\mathrm{eff}}}{N\chi_{\mathrm{eff}}})^{2}}}\bigg|_{Z(0)}^{Z(t_{\rm opt})}\\
        &\approx -\frac{f(Z)}{\sqrt{1-(\frac{\delta_{\mathrm{eff}}}{N\chi_{\mathrm{eff}}})^{2}}} \bigg|_{\sqrt{1-\frac{1}{2N}}}^{\frac{2\delta_{\mathrm{eff}}}{N\chi_{\mathrm{eff}}}-1}\\
        &\approx\frac{\ln\left\{\frac{16\left[1-(\frac{\delta_{\mathrm{eff}}}{N\chi_{\mathrm{eff}}})^{2}\right]}{1-\sqrt{1-(\frac{\delta_{\mathrm{eff}}}{N\chi_{\mathrm{eff}}})^{2}}}\right\}-2\arctanh\left[\frac{1-\frac{\delta_{\mathrm{eff}}}{N\chi_{\mathrm{eff}}}}{\sqrt{1-(\frac{\delta_{\mathrm{eff}}}{N\chi_{\mathrm{eff}}})^{2}}}\right]+\ln{N}}{2\sqrt{1-(\frac{\delta_{\mathrm{eff}}}{N\chi_{\mathrm{eff}}})^{2}}},
    \end{aligned}
    \end{equation}
    with 
    \begin{equation}
    \begin{aligned}
    f(Z) = \arctanh\left[\frac{1-Z+\sqrt{\left(1+Z\right)^{2}-(\frac{2\delta_{\mathrm{eff}}}{N\chi_{\mathrm{eff}}})^2}}{2\sqrt{1-(\frac{\delta_{\mathrm{eff}}}{N\chi_{\mathrm{eff}}})^{2}}}\right].
    \end{aligned}
    \end{equation}

    This result follows from the properties of the uncertainty patch. The lower bound, $\sqrt{1-\frac{1}{2N}}$, of the integral originates from the separatrix of the uncertainty patch, which sets the initial stretching position, while the upper bound, $\frac{2\delta_{\mathrm{eff}}}{N\chi_{\mathrm{eff}}}-1$, corresponds to the critical point of variance change on the separatrix. The Taylor expansion used above is
    \begin{equation}
    \begin{aligned}
    \arctanh\left[\frac{1+\sqrt{\left(1+\sqrt{1-\frac{1}{2N}}\right)^2-4(\frac{\delta_{\mathrm{eff}}}{N\chi_{\mathrm{eff}}})^2}-\sqrt{1-\frac{1}{2N}}}{2\sqrt{1-(\frac{\delta_{\mathrm{eff}}}{N\chi_{\mathrm{eff}}})^2}}\right]  = \frac{1}{2} \ln\left\{\frac{16\left[1-(\frac{\delta_{\mathrm{eff}}}{N\chi_{\mathrm{eff}}})^{2}\right]}{1-\sqrt{1-(\frac{\delta_{\mathrm{eff}}}{N\chi_{\mathrm{eff}}})^{2}}}\right\}+\frac{1}{2} \ln{N}+\mathcal{O}\left(\frac{1}{N}\right).
    \end{aligned}
    \end{equation}
    When $\frac{\delta_{\mathrm{eff}}}{N\chi_{\mathrm{eff}}}=\sqrt{2}-1$, $N\chi_{\mathrm{eff}} t_{\rm opt} \simeq {1.9+0.55\ln{N}}$, therefore 
    \begin{equation}
    \begin{aligned}
    {\chi_{\rm eff}} t_{\rm opt} \simeq \frac{1.9+0.55\ln{N}}{N}. \label{time_scaling}
    \end{aligned}
    \end{equation}
    When $0 \leq t \leq t_{\rm opt}$, we can determine $Z(t)$ by solving the unique root of the following equation:
    \begin{equation}
    \begin{aligned}
    f[Z(t)] - f[Z(0)] + \sqrt{{(N\chi_{\mathrm{eff}}})^{2}-\delta_{\mathrm{eff}}^{2}} t =0.
    \end{aligned}
    \end{equation}
    Due to the special symmetry of the phase space of the TAT-and-turn Hamiltonian, the same result can be obtained at another saddle point $(0, 0, -1)$ for the similar calculation.

    \section{III. Heisenberg scaling and comparison with spin cat states}
    In the main text, we have shown that the GHZ-like states generated by Floquet-engineered TAT-and-turn dynamics exhibit metrological properties that follow Heisenberg scaling, similar to the spin cat states. 
    In this section, we further analyze the QFI scaling of the GHZ-like states, assuming the generated GHZ-like states can be approximated as spin cat states.
    Compared to the semi-classical approach, which estimates the critical point $Z(t_{\rm opt}) \approx \frac{2\delta_{\mathrm{eff}}}{N\chi_{\mathrm{eff}}}-1$ through Eq.~\eqref{qfi_c}, the results obtained here are closer to those from a full quantum treatment.

    Applying the time in Eq.~\eqref{time_scaling} and the critical detuning in Eq.~\eqref{detuning_c}, the GHZ-like state $\lvert \psi \rangle_{\rm GHZ-like}$ whose two main parts are symmetrically distributed near the pole of $y$-axis can be generated by Floquet-engineered TAT-and-turn dynamics starting from the initial state $\lvert \uparrow \rangle^{\otimes N}$ with $\alpha = 0$. To align its orientation with a spin-cat state consisting of two SCSs with the same azimuthal angle $\varphi$ and polar angle symmetric about $\vartheta = \pi/2$, the GHZ-like state $\lvert \psi \rangle_{\rm GHZ-like}$ needs to be rotated to the $z$-axis via $e^{-i\hat{J}_x \pi/2}$. The expectation of $\hat{J}_z$ for the GHZ-like state $\langle \hat{J}_z \rangle_{\rm GHZ-like}^{'} = 0$ with $\lvert \psi \rangle_{\rm GHZ-like}^{'} = e^{-i\hat{J}_x \pi/2} \lvert \psi \rangle_{\rm GHZ-like}$, and the probability distribution is mainly concentrated around the two ends of the $z$-axis.
    Thus the symmetrical GHZ-like entangle state can be approximately regarded as a macroscopic superposition spin coherent state like a spin cat state~\cite{PhysRevA.98.012129}, which is given by 
    \begin{equation}
    \lvert \psi \rangle_{\rm GHZ-like}^{'} \sim |\Psi(\vartheta)\rangle_{\mathrm{CAT}} \approx \frac{1}{\sqrt{2}}(|\vartheta,\varphi\rangle_{\rm SCS}+|\pi-\vartheta,\varphi\rangle_{\rm SCS}),
    \end{equation}
    where $\vartheta = \arccos \overline{Z}$ and the normalized average spin length $\overline{Z}=\frac{2}{N}\sum_{m=-N/2}^{N/2} \vert m \vert P_m =\frac{2}{N}\sum_{m=-N/2}^{N/2} \vert m \vert|\langle m|\psi\rangle^{'}_{\rm{GHZ-like}}|^2$ with the eigenstate $|m \rangle$ of $\hat{J}_z$. This expression approximates the GHZ-like state as a spin cat state with the same normalized average spin length is $\overline{Z}$.
    The probability distributions of both the GHZ-like state and its corresponding spin cat state are shown in Fig.~\ref{cat}(a).
    In this case, the QFI of this GHZ-like state is approximately written as
    \begin{equation}
    F_Q^{\rm GHZ-like} \sim F_Q^{\rm CAT} \approx N^2 \cos^2\vartheta = N^2{\overline{Z}}^2.
    \label{QFI_cat}
    \end{equation}

    Further, we use Eq.~\eqref{QFI_cat} to calculate the QFI of GHZ-like states generated by TAT-and-turn dynamics for different particle number $N$. As shown in Fig.~\ref{cat}(b), the results are in agreement with the quantum results, illustrating that the best achievable QFI of Floquet-engineered TAT-and-turn exhibits the Heisenberg-limited scaling $F_Q^{\rm max} \propto N^2$, which is consistent with the QCRB of a spin cat state.
    \begin{figure}[hbt]
        \centering
        \includegraphics[width=0.88\linewidth]{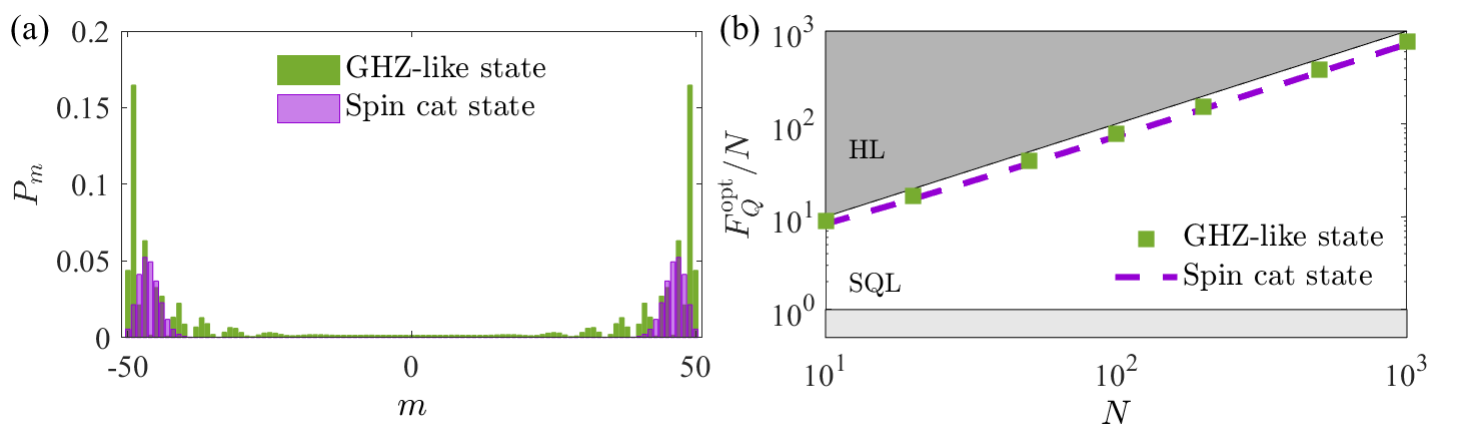} 
        \caption
        {(a) The probability distributions of GHZ-like state generated by Floquet-engineered TAT-and-turn dynamics and its corresponding spin cat state with $\theta=\arccos{\overline{Z}}$ for particle number $N=100$. (b) The QFI of GHZ-like states generated by Floquet-engineered TAT-and-turn dynamics and their corresponding spin cat states with $\theta=\arccos{\overline{Z}}$ versus different particle number $N$.
        }
        \label{cat}
    \end{figure}

    \section{IV. Interaction-based readout}
    The interaction-based readout (IBR) offers a potent protocol for attaining high-precision Heisenberg-limited measurements. When implementing IBR for the final state following the given dynamics evolution, the achievable measurement precision depends on both the directions for encoding the signal into the probe state and for detecting the final state.
    The signal encoding operator $\hat{J}_{\vec{n}}$ determines the QFI of a given probe state, thus setting a bound (i.e., QCRB) on the ultimate measurement precision.
    The detection using observable operator $\hat{J}_{\vec{m}}$ determines how closely we can approach the QCRB from a given measurement direction. Optimizing both directions $\vec{n}$ and $\vec{m}$ is crucial for improving measurement precision. In the following, we introduce the method to optimize the measurement direction with a given sensing direction~\cite{Schulte2020ramsey, PhysRevA.107.052613}.

    In the general process of IBR, for an arbitrary observable operator $\hat{J}_{\vec{m}}$, we transform the Schrödinger picture into the Heisenberg picture, i.e.
    \begin{equation}
    \begin{aligned}
    \hat{\tilde{J}}_{\vec{m}} &= \hat{U}^\dagger_1 \hat{R}_{\vec{n}}^\dagger(\phi) \hat{U}^\dagger_2 \hat{J}_{\vec{m}} \hat{U}_2 \hat{R}_{\vec{n}}(\phi) \hat{U}_1 \\
    &= \hat{U}^\dagger_1 e^{i\phi \hat{J}_{\vec{n}} } \hat{U}_1 \hat{U}^\dagger_1 \hat{U}^\dagger_2 \hat{J}_{\vec{m}} \hat{U}_2 \hat{U}_1 \hat{U}^\dagger_1 e^{-i \phi \hat{J}_{\vec{n}}} \hat{U}_1 \\
    & =  e^{i\phi \hat{U}^\dagger_1 \hat{J}_{\vec{n}} \hat{U}_1 }  \hat{U}^\dagger_1 \hat{U}^\dagger_2 \hat{J}_{\vec{m}} \hat{U}_2 \hat{U}_1  e^{-i \phi \hat{U}^\dagger_1 \hat{J}_{\vec{n}} \hat{U}_1} ,
    \end{aligned}
    \end{equation}
    where $\hat{R}_{\vec{n}}(\phi) = e^{-i\hat{J}_{\vec{n}} \phi}$ for encoding signal, $\hat{U}_1= e^{-i\hat{H}_1 t_1}$ for generating entangled states, and $\hat{U}_2= e^{-i\hat{H}_2 t_2}$ for signal amplification through disentanglement, typically time-reversal IBR with $\hat{H}_2=-\hat{H}_1$ and $t_2 = t_1$. When half-population difference readout is performed directly instead of IBR, the Hamiltonian $\hat{H}_2 = 0$.
    Therefore, the measurement response can be calculated and expressed in matrix form as
    \begin{equation}
    \frac{\partial \langle \hat{\tilde{J}}_{\vec{m}} \rangle }{\partial\phi} = i \langle [ \hat{U}^\dagger_1 \hat{J}_{\vec{n}} \hat{U}_1, \hat{\tilde{J}}_{\vec{m}} ] \rangle = \vec{n}^T \mathcal{M} \vec{m},
    \end{equation}
    where $\mathcal{M}$ is a $3 \times 3$ matrix with elements $\mathcal{M}_{\mu \nu} = i\langle [ \hat{U}^\dagger_1 \hat{J}_{\mu} \hat{U}_1, \hat{\tilde{J}}_{\nu}] \rangle( \mu, \nu= x, y, z)$. Similarly, the measurement uncertainty can be expressed as
    \begin{equation}
    (\Delta\hat{\tilde{J}}_{\vec{m}})^2 = \langle \hat{\tilde{J}}_{\vec{m}}^2 \rangle - \langle \hat{\tilde{J}}_{\vec{m}} \rangle^2 = {\vec{m}}^T \mathcal{Q} {\vec{m}},
    \end{equation}
    where $\mathcal{Q}$ represents the $3 \times 3$ covariance matrix with elements $\mathcal{Q}_{\mu \nu} = \langle \{\hat{\tilde{J}}_{\mu}, \hat{\tilde{J}}_{\nu}\}  \rangle / 2 - \langle \hat{\tilde{J}}_{\mu} \rangle \langle \hat{\tilde{J}}_{\nu} \rangle$. Using the error propagation formula, the achievable measurement sensitivity can be expressed as
    \begin{equation}
    [\Delta\phi(\vec{n},\vec{m})]^{-2} = \frac{|\partial\langle\hat{\tilde{J}}_{\vec{m}} \rangle/\partial\phi|^2}{(\Delta\hat{\tilde{J}}_{\vec{m}})^2}=\frac{(\vec{n}^T \mathcal{M} \vec{m})^2}{\vec{m}^T \mathcal{Q} \vec{m}}. \label{epf}
    \end{equation}

    To find the optimal measurement direction, one can apply the Cauchy-Schwarz inequality $(\vec{u}^T\vec{v})^2 \leq \vec{u}^T\vec{u} \vec{v}^T\vec{v}$, which implies that the measurement sensitivity satisfies
    \begin{equation}
    [\Delta\phi(\vec{n},\vec{m})]^{-2}  \leq \mathop{\max}_{\vec{m}}[\Delta\phi(\vec{n}, \vec{m})]^{-2}  = \vec{n}^T \mathcal{K} \vec{n}, \label{epf1}
    \end{equation}
    in which $\mathcal{K} = \mathcal{M} \mathcal{Q}^{-1} \mathcal{M}^T$ depends on the selection of $\vec{n}$. The equality is achieved at $\vec{m}_{\rm opt} = \mathcal{N}_c \mathcal{Q}^{-1} \mathcal{M}^T \vec{n}$ with $\mathcal{N}_c$ the normalization
    constant. This applies to a given sensing operator $\hat{J}_{\vec{n}}$ in the general case, including both $\phi=0$ and $\phi \neq 0$.

    Thus, by applying the optimal detection direction, i.e. $\vec{m}_{\rm opt} = \mathcal{N}_c \mathcal{Q}^{-1} \mathcal{M}^T \vec{n}$, the measurement precision is only determined by the encoding signal direction $\vec{n}$.
    In this work, we choose the sensing direction $\vec{n}_{\rm max}$ that corresponds to the maximum QFI value of the entangled state to encode the signal.
    By combining Eq.~\eqref{epf} and Eq.~\eqref{epf1}, we obtain
    \begin{equation}
    \mathop{\max}_{\vec{m}}[\Delta\phi(\vec{n}_{\rm max}, \vec{m})]^{-2}  = \vec{n}_{\rm max}^T \mathcal{K} \vec{n}_{\rm max}.
    \end{equation}
    Note that the phase $\phi$ cannot be equal to $0$, as this would cause the matrix $\mathcal{K}$ will lose the direction $\vec{n}_{\rm max}$ and become restricted to the plane perpendicular to the polarization direction.

    In real experiments or application scenarios, both the sensing and measurement operators will select $\hat{J}_z$, the former for free evolution (phase accumulation) and the latter for half-population readout. We can equivalently perform the desired $\hat{J}_{\vec{n}}$ and $\hat{J}_{\vec{m}}$ with the pulses in the $x-y$ plane. To achieve an equivalent generator $\hat{J}_{\vec{n}}$, assuming the pulses are along $x$-axis and $y$-axis respectively, four pulses are applied before and after the free evolution. Thus we have
    \begin{equation}
    \begin{aligned}
    e^{i\vartheta_n \hat{J}_{x}} e^{i \varphi_n \hat{J}_{y}} e^{-i\phi\hat{J}_{z}} e^{-i\varphi_n \hat{J}_{y}} e^{-i \vartheta_n \hat{J}_{x}} &= e^{i\vartheta_n \hat{J}_{x}}  e^{-i\phi(\hat{J}_{z}\cos\varphi_n - \hat{J}_{x}\sin\varphi_n)}  e^{-i \vartheta_n \hat{J}_{x}} \\
    &= e^{-i\phi[(\hat{J}_{z}\cos\vartheta_n + \hat{J}_{y}\sin\vartheta_n )\cos\varphi_n - \hat{J}_{x}\sin\varphi_n]} \\
    &= e^{-i\phi( - \hat{J}_{x}\sin\varphi_n + \hat{J}_{y} \cos\vartheta_n \sin\vartheta_n + \hat{J}_{z}\cos\varphi_n \cos\vartheta_n )}.
    \end{aligned}
    \end{equation}
    When setting $\vartheta_n = \arctan{(n_y/n_z)}$, $\varphi_n = -\arcsin{n_x}$, the above formula is equivalent to $e^{-i\phi\hat{J}_{\vec{n}}} = e^{-i\phi( n_x \hat{J}_{x} + n_y \hat{J}_{y}  + n_z \hat{J}_{z})}$. Similarly, for half-population readout, one requires two pulses applied to the final state, and we can obtain the following equation
    \begin{equation}
    \begin{aligned}
    \langle \hat{J}_{\vec{m}} \rangle_f = \langle e^{i\vartheta_m \hat{J}_{x}} e^{i \varphi_m \hat{J}_{y}} \hat{J}_z e^{-i\varphi_m \hat{J}_{y}} e^{-i \vartheta_m \hat{J}_{x}} \rangle_f,
    \end{aligned}
    \end{equation}
    where $\vartheta_m = \arctan{(m_y/m_z)}$, $\varphi_m = -\arcsin{m_x}$.
    And the required sensing direction $\vec{n}$ can be obtained by constructing QFIM, the corresponding optimal measurement direction $\vec{m}_{\rm opt} = \mathcal{N}_c \mathcal{Q}^{-1} \mathcal{M}^T \vec{n}$ in time-reversal IBR.

    As shown in Fig.~\ref{fig_appdix2}~(a), we analyze the measurement precision of TAT-and-turn Hamiltonian under time-reversal IBR using above method. The results show that the method we proposed is close to the QCRB with excellent performance of measurement precision. The entangled state generated in a short time is a spin squeezed state, and after a longer time, the generated state becomes a non-Gaussian state. During the process, the measurement precision can approach the QCRB. From Fig.~\ref{fig_appdix2}~(b), we find that the measurement precision gradually converges as $\phi$ approaches $0$, which is conducive to high-precision measurement of tiny signals. 

    \begin{figure}[ht]
        \centering
        \includegraphics[width=0.88\linewidth]{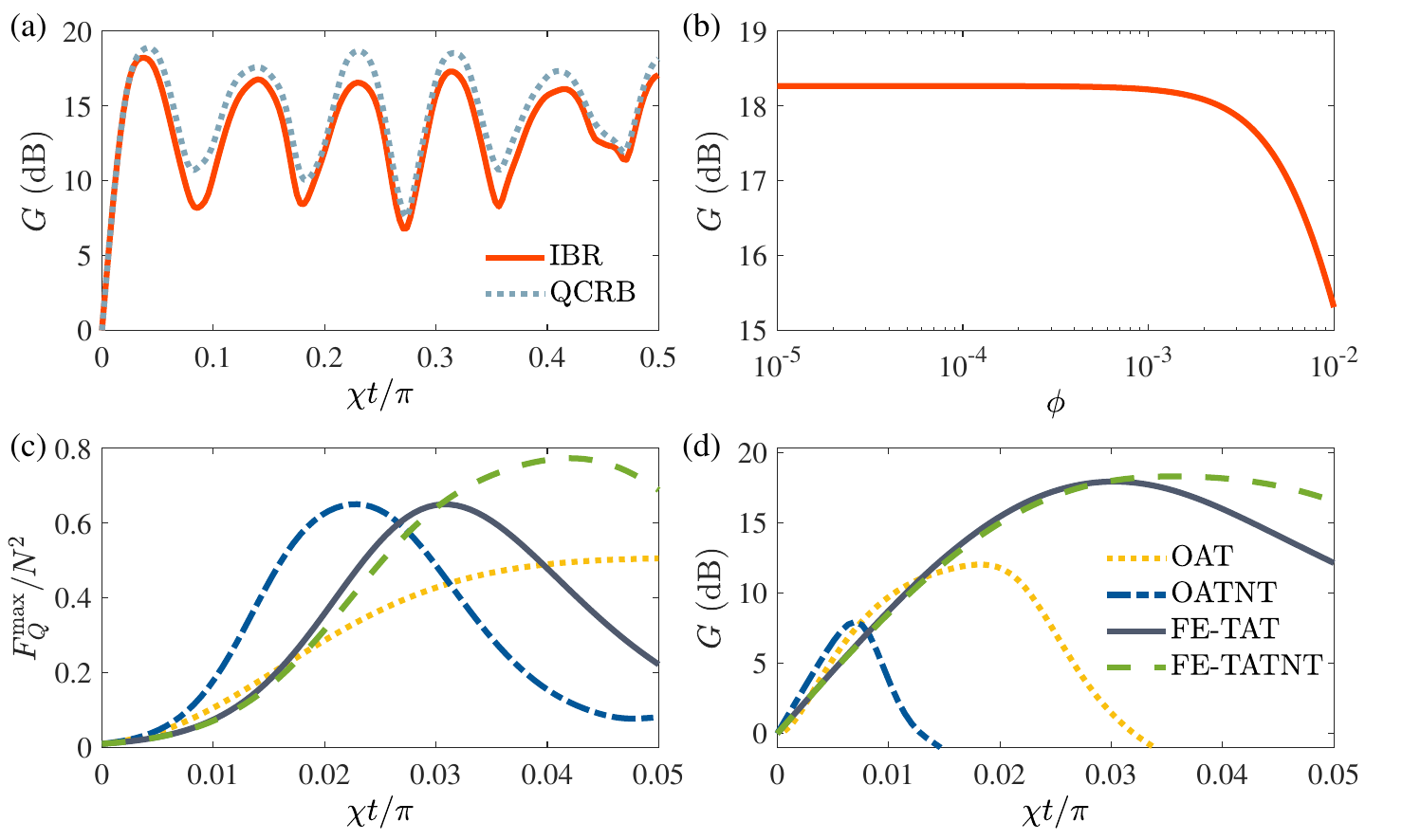} 
        \caption
        {(a) The IBR and QCRB for TAT-and-turn. (b) The attainable measurement precision of TAT-and-turn at $t=0.12$ via IBR for $\phi$ approaching 0. (c) The time evolution of maximum QFI for OAT, OAT-and-turn, Floquet-engineered TAT, and Floquet-engineered TAT-and-turn. (d) The IBR without flipping the sign of nonlinear interaction for OAT, OAT-and-turn, Floquet-engineered TAT and Floquet-engineered TAT-and-turn. The metrological gain is defined as $G=20 \log _{10} \left[{(\Delta \phi)_{\mathrm{SQL}}}/{\Delta \phi}\right]$ with $(\Delta\phi)_{\mathrm{SQL}}=1/{\sqrt{N}}$ denoting the measurement precision of SQL. The phase $\phi=1/1000$ and particles number $N=100$ are chosen for simulation.}
        \label{fig_appdix2}
    \end{figure}

    \section{V. Different dynamics with continuous Floquet engineering}
    Varying system parameters, the Floquet-engineered TAT-and-turn~\eqref{TATNT Hamiltonian} dynamics can be reduced to typical TAT, OAT-and-turn, or OAT dynamics.
    When $\delta=0$, the effective TAT-and-turn Hamiltonian can be reduced to an effective TAT Hamiltonian $\hat{H}_{\rm TAT}=\chi_{\rm eff}( \hat{J}_{\beta}^{2} - \hat{J}_{\alpha}^{2})$~\cite{PhysRevA.92.013623,PhysRevLett.129.090403}. 
    Further, setting $\Omega(t)=\Omega_{0}$ as a constant and $\delta=0$, the Hamiltonian~\eqref{H_FE} is reduced to an OAT-and-turn Hamiltonian $\hat{H}_\mathrm{OATNT} = \chi \hat{J}_{z}^{2} + \Omega_{0}\hat{J}_{\alpha}$~\cite{PhysRevA.67.013607,PhysRevA.92.023603}. 
    If $\delta=0$ and $\Omega(t)=0$, the Hamiltonian~\eqref{H_FE} becomes an OAT Hamiltonian $\hat{H}_\mathrm{OAT}=\chi \hat{J}_{z}^{2}$.
    In following, we firstly investigate the QFI of states generated by OAT, OAT-and-turn, Floquet-engineered TAT, and Floquet-engineered TAT-and-turn dynamics and subsequently study the IBR without flipping the sign of nonlinear interaction using OAT, OAT-and-turn, Floquet-engineered TAT, and Floquet-engineered TAT-and-turn dynamics.

    As depicted in Fig.~\ref{fig_appdix2}~(c), we present the time evolution of the maximum QFI of states generated by OAT, OAT-and-turn, Floquet-engineered TAT, and Floquet-engineered TAT-and-turn dynamics.
    The results show that the state generated by OAT-and-turn dynamics can grow QFI at the fastest rate in a short period of time and the state produced by Floquet-engineered TAT-and-turn dynamics can attain the highest QFI during the considered period.
    Furthermore, by comparing OAT with OAT-and-turn and Floquet-engineered TAT with Floquet-engineered TAT-and-turn, we discover that the turn dynamics can facilitate the preparation of states with higher QFI.

    The IBR provides a powerful protocol for achieving high-precision Heisenberg-limited measurements, however, it requires to reverse the dynamics of an interacting many-body quantum system by inverting the sign of nonlinear interaction which is still experimentally challenging in some systems.
    Here, we concentrate on and explore the IBR without reversing the sign of nonlinear interaction through OAT, OAT-and-turn, Floquet-engineered TAT, and Floquet-engineered TAT-and-turn dynamics.
    For OAT dynamics, the implementation of IBR without flipping the sign of Hamiltonian is written as $\lvert\psi_{f}\rangle=e^{-i \hat{H}_{\mathrm{OAT}} t}\hat{R}_n(\phi)e^{-i \hat{J}_x (\pi-2\gamma_\mathrm{opt})}e^{-i \hat{H}_{\mathrm{OAT}} t}\lvert\psi_{i}\rangle$,
    where the initial state $\lvert\psi_i\rangle$ is spin coherent state polarized along $x$, the phase is encoded by rotation $\hat{R}_n(\phi)=e^{-i \hat{J}_y \phi}$ along $y$-axis and $\gamma_\mathrm{opt}=\frac{1}{2}\arctan\left(B/A\right)$ is the optimal angle of squeezing with ${A}=1-\cos^{N-2}\left(2\chi t\right)$ and ${B}=4\cos^{N-2}\left(\chi t\right)\sin\left(\chi t\right)$~\cite{PhysRevA.110.022407}.
    For OAT-and-turn dynamics, the implementation of IBR without flipping the sign of Hamiltonian can be expressed as $\lvert\psi_{f}\rangle=e^{-i \hat{H}_{\mathrm{OATNT}} t}\hat{R}_n(\phi)e^{-i \hat{J}_x \pi/2}e^{-i \hat{H}_{\mathrm{OATNT}} t}\lvert\psi_{i}\rangle$, where the initial state $\lvert\psi_i\rangle$ is chosen as spin coherent state polarized along $x$ and the phase is encoded by rotation $\hat{R}_n(\phi)=e^{-i\phi (\hat{J}_z-\hat{J}_y)/\sqrt{2} }$.
    For realizing time-reversal IBR without flipping the nonlinear interaction of Floquet-engineered TAT-and-turn, it only needs to adjust the phase of external coupling field and flip the detuning, as discussed in main text.
    In addition, the time-reversal IBR without flipping the nonlinear interaction of Floquet-engineered TAT is the case of Floquet-engineered TAT-and-turn with detuning $\delta=0$.
    We show the results of the IBR for OAT, OAT-and-turn, Floquet-engineered TAT and Floquet-engineered TAT-and-turn without flipping the sign of nonlinear interaction in Fig.~\ref{fig_appdix2}~(d).
    In a short period of time, the OAT-and-turn can achieve the highest metrological gain, and the Floquet-engineered TAT-and-turn protocol has the best performance in the considered time.

    \section{VI. Effect of variations in driving frequency and Rabi frequency}
    When the driving frequency is high and the suitable Rabi frequency $\Omega_{0}\simeq1.6262\omega$, the FE Hamiltonian is transformed into the ideal TAT-and-turn Hamiltonian. 
    In this section, we explore the dependence of these two crucial parameters.
    \begin{figure}[hbt]
        \centering
        \includegraphics[width=0.88\linewidth]{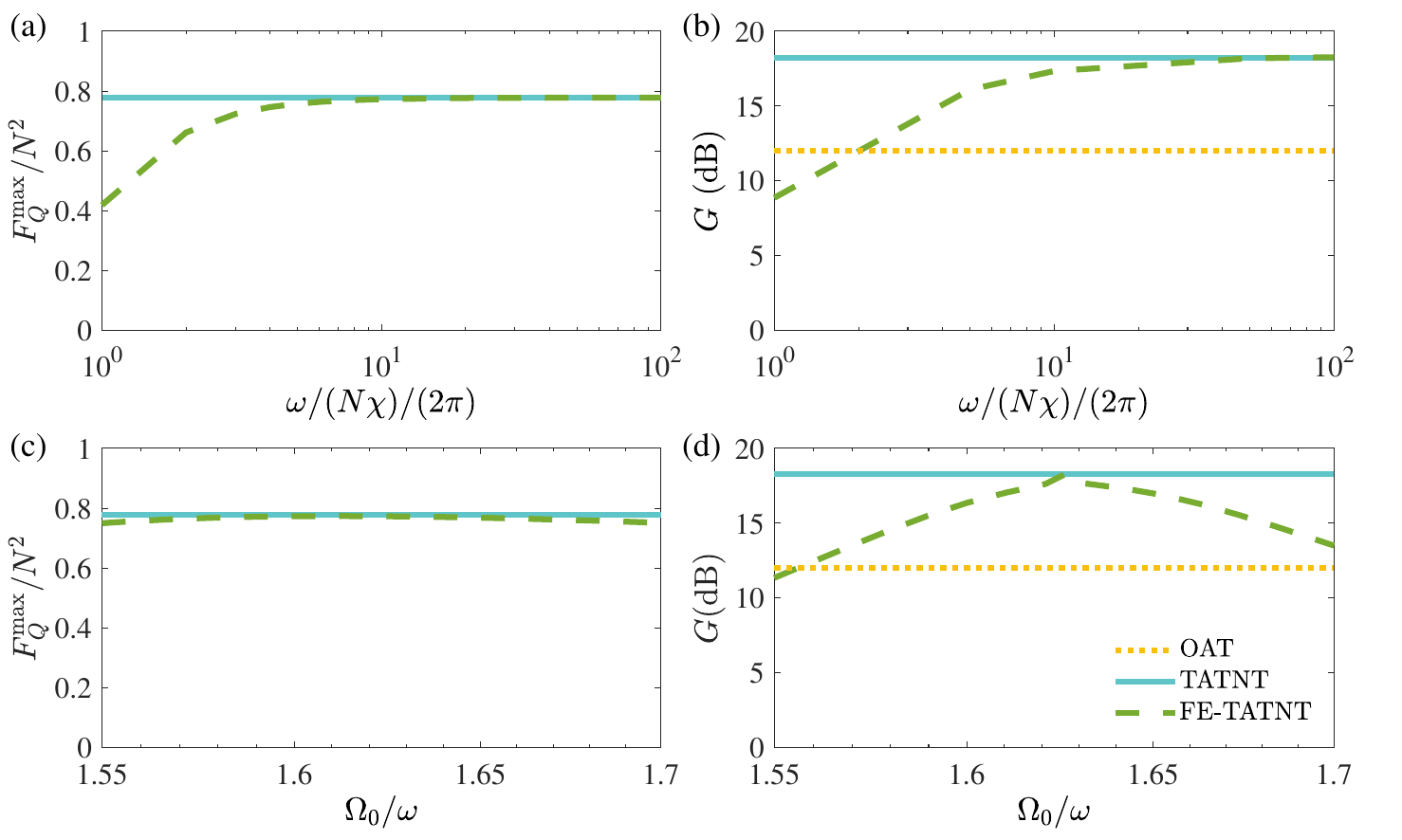} 
        \caption
        {The dependence of driving frequency for (a) state preparation and (b) IBR.
        The dependence of imperfect Rabi frequency for (c) state preparation and (d) IBR. The best performance of IBR for OAT with fixed nonlinear interaction in Fig.~\ref{fig_appdix2}(d) is shown for comparison and the result of ideal TATNT is also shown for verifying the perfect condition. The phase $\phi=1/1000$ and particles number $N=100$ are chosen for simulation.
        }
        \label{SM_ferquency}
    \end{figure}

    Firstly, we examine the influence of the driving frequency on state preparation and IBR.
    As depicted in Figs.~\ref{SM_ferquency}~(a) and (b), with the increase of the driving frequency, the QFI or metrological gain also rises.
    For state preparation, the QFI generated through the Floquet-engineered TAT-and-turn approaches that via the ideal TAT-and-turn when the driving frequency is $10N\chi$ for a particle number $N=100$, as shown in Fig.~\ref{SM_ferquency}~(a).
    For IBR, the metrological gain generated through the Floquet-engineered TAT-and-turn approaches that via the ideal TAT-and-turn when the driving frequency is $10^2N\chi$ for a particle number $N=100$, as shown in Fig.~\ref{SM_ferquency}~(b).
    The above results are consistent with our previous analysis that the FE Hamiltonian is equivalent to the TAT-and-turn Hamiltonian when the driving frequency is sufficiently large ($\omega \gg N\chi$).

    Subsequently, we investigate the effect of the variation of the Rabi frequency.
    The state generated by the Floquet-engineered Hamiltonian under different ratios $\Omega_{0}/\omega$ has significant QFI, as shown in Fig.~\ref{SM_ferquency}~(c).
    In addition, compared with the best performance of IBR for OAT with fixed nonlinear interaction, the Floquet-engineered TAT-and-turn protocol can achieve a greater metrological gain under different ratios $\Omega_{0}/\omega$, as shown in Fig.~\ref{SM_ferquency}~(d), indicating that the Floquet-engineered TAT-and-turn is robust against the variation of the Rabi frequency.

    \section{VII. Robustness against decoherence}
    In the absence of decoherence, the GHZ (GHZ-like) states can be ideally generated through OAT, XYZ~\cite{PhysRevLett.132.113402} or Floquet-engineered TATNT dynamics, achieving a precision approaching the Heisenberg limit.
    However, the existence of noise will destroy the coherence of the state and thereby reduce the achievable precision in practice.
    To analyze the influence of decoherence in generating GHZ-like states, we separately consider two decoherence effects here. One is the spontaneous emission~\cite{PhysRevA.4.1791} that can be described by the Lindblad operator ${\hat{L}_{1}}=\sqrt{\Gamma/2}\hat{J}_{-}$ with $\hat{J}_{-}=\hat{J}_{x}-i\hat{J}_{y}$ and the spontaneous emission rate $\Gamma$. The other is the correlated dephasing~\cite{Dorner_2012,PhysRevA.88.062113} that can be described by the Lindblad operator ${\hat{L}_{2}}=\sqrt{\gamma/2}\hat{J}_{z}$ with the dephasing rate $\gamma$.
    The master equation is given by 
    \begin{equation}
        \dot{\hat{\rho}}=-i[\hat{H},\hat{\rho}]+\left(\hat{L}_{i} \hat{\rho} \hat{L}_{i}^\dagger-\frac{1}{2}\hat{L}_{i}^\dagger \hat{L}_{i}\hat{\rho}-\frac{1}{2}\hat{\rho} \hat{L}_{i}^\dagger \hat{L}_{i}\right),
    \end{equation}
    where the $\hat{\rho}$ represents the density matrix, $\hat{H}$ is the Hamiltonian of the system and $\hat{L}_{i}$ is selected as $\hat{L}_{1}$ or $\hat{L}_{2}$ for spontaneous emission or correlated dephasing.

    \begin{figure}[hbt]
        \centering
        \includegraphics[width=1\linewidth]{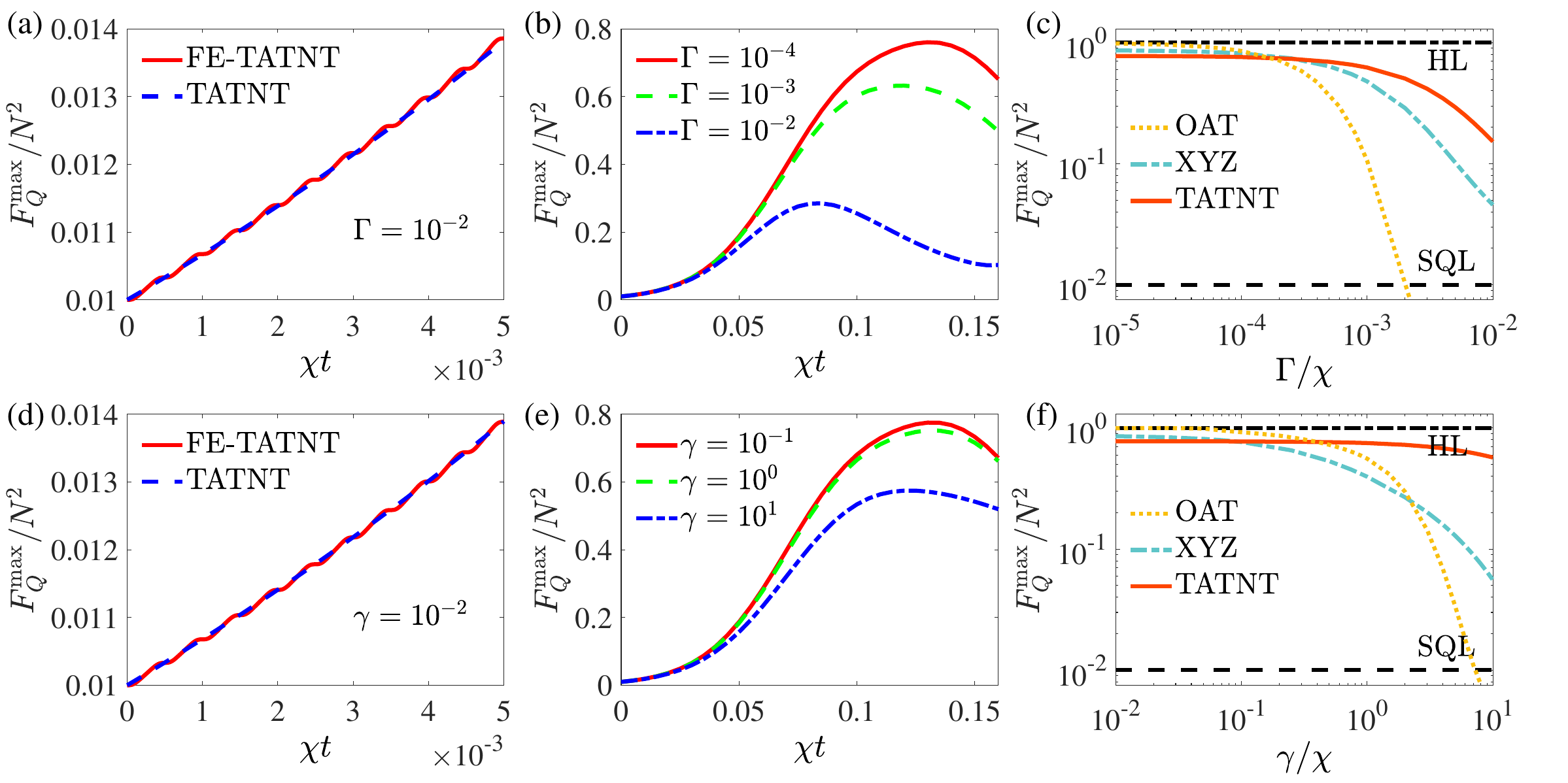} 
        \caption
        {Comparison of time evolution of Floquet-engineered TATNT and ideal TATNT under (a) spontaneous emission or (d) correlated dephasing . 
        The time evolution of TATNT under (b) different spontaneous emission rate $\Gamma$ or (e) different dephasing rate $\gamma$. Robustness against (c) spontaneous emission or (f) correlated dephasing for generating GHZ (GHZ-like) state by OAT, XYZ or TATNT dynamics. The particles number $N=100$ and driving frequency $\omega=2\pi\times10N\chi$ are chosen for simulation. 
        }
        \label{decoherence_SM}
    \end{figure}

    We first investigate the influence of spontaneous emission on the system. We verify the time evolution of QFI produced by Floquet-engineered TATNT and ideal TATNT under spontaneous emission over a short period of time, as shown in Fig.~\ref{decoherence_SM}~(a). The results of Floquet-engineered TATNT and ideal TATNT are highly consistent.
    Since the numerical computation of time-dependent FE-TATNT evolution over a long period is extremely time-consuming, we will demonstrate the robustness to decoherence using ideal TATNT in the main text and hereinafter.
    As shown in Fig.~\ref{decoherence_SM}~(b), we present the long-term evolution of the maximum QFI of states generated by TATNT under spontaneous emission, with three different spontaneous emission rates,  $\Gamma=10^{-4},10^{-3},10^{-2}$.
    The results reveal that the QFI of the prepared state decreases as the spontaneous emission rate increases. 
    Furthermore, we investigate the generation of GHZ (GHZ-like) state by OAT, XYZ or TATNT under spontaneous emission.
    As depicted in Fig.~\ref{decoherence_SM}~(c), the generation of GHZ-like state by TATNT under spontaneous emission shows the best performance, demonstrating its excellent robustness against decoherence induced by spontaneous emission.

    Subsequently, we study the impact of decoherence due to correlated dephasing.
    We analyze the time evolution of QFI produced by Floquet-engineered TATNT and ideal TATNT under correlated dephasing over a short period of time, as shown in Fig.~\ref{decoherence_SM}~(d). Once again, the results for both approaches are highly consistent.
    As depicted in Fig.~\ref{decoherence_SM}~(e), we present the long-term evolution of the maximum QFI of states generated by TATNT under correlated dephasing, for three different dephasing rates, $\gamma=10^{-1},10^{0},10^{1}$.
    It is observed that the QFI of the prepared state decreases as the dephasing rate increases.
    Additionally, we investigate the generation of GHZ (GHZ-like) state by OAT, XYZ or TATNT under correlated dephasing.
    As shown in Fig.~\ref{decoherence_SM}~(f), the generation of GHZ-like state by TATNT under correlated dephasing performs best, further highlighting the exceptional robustness of TATNT against decoherence due to correlated dephasing.
\end{widetext}
\end{document}